\documentclass[lettersize,journal]{IEEEtran}

\usepackage{amsmath,amsfonts}
\usepackage{algorithmic}
\usepackage{array}
\usepackage{pifont}
\usepackage{textcomp}
\usepackage{stfloats}
\usepackage{url}
\usepackage{color}
\usepackage{verbatim}
\usepackage{graphicx}
\usepackage{subfigure} 
\usepackage{soul}
\usepackage{booktabs}
\usepackage{makecell}
\usepackage{caption}

\def\BibTeX{{\rm B\kern-.05em{\sc i\kern-.025em b}\kern-.08em
    T\kern-.1667em\lower.7ex\hbox{E}\kern-.125emX}}
\usepackage{balance}

\usepackage{multirow}

\usepackage{pifont}
\usepackage{float}

\begin{document}
\title{An Argument-Principle Based Stability Assessment Method for Grey-Box DFIG Systems}

\author{Tao Zhang, \emph{ Student Member}, \emph{IEEE}, Zhiguo Hao, \emph{ Member}, \emph{ IEEE}, Songhao Yang, \emph{ Member}, \emph{ IEEE}, \\Hongyue Ma, Baohui Zhang, \emph{Fellow}, \emph{IEEE}
\thanks{This work was supported by National Natural Science Foundation of China (52007143), China Postdoctoral Science Foundation (2021M692526) and Key R \& D Program of Shaanxi Province(2022GXLH-01-06).

Tao Zhang, Zhiguo Hao, Songhao Yang, Hongyue Ma, Baohui Zhang are with State Key Laboratory of Electrical Insulation and Power Equipment, Xi'an Jiaotong University, Xi'an, China (e-mail: {songhaoyang, zhghao}@xjtu.edu.cn).}
}
\markboth{Journal of IEEE,~Vol.~X, No.~X, XXX~XXX}%
{How to Use the IEEEtran \LaTeX \ Templates}

\maketitle
\begin{abstract}
Considerable efforts have been made to analyze the small-signal stability of doubly fed induction generator (DFIG) systems. However, commercial confidentiality and frequency coupling make the DFIG system a grey-box multiple-input-multiple-output (MIMO) system with highly challenging stability analysis. This paper proposes an Argument-principle based stability assessment method to analyze the stability of the grey-box DFIG system. The frequency sweeping technique is first used to acquire the MIMO model of the black-box device, as well as the determinant of the system’s return difference matrix. Then a stability criterion based on the determinant trajectory is presented. This criterion applies to the stability analysis of grey-box MIMO systems without detailed system models. Further, a critical-pole estimation method with trajectory information is developed to assess the dominant mode of the target system. The simulation and hardware-in-loop experiment results demonstrate the effectiveness of the proposed method. Finally, some concerns about this method, such as  model selection, estimation errors and application potential, are thoroughly analyzed and clarified.
	
\end{abstract}


\begin{table}[H]
\normalsize
	\renewcommand\arraystretch{1.05}
	\tabcolsep=0.4cm
	\caption*{A\textsc{cronyms}}
				\begin{tabular}{ l  l  }
			DFIG & doubly fed induction generator.\\
			SISO & single-input single-output. \\
			MIMO  & Multiple-input multiple-output. \\
			GNC & generalized Nyquist criterion.  \\
			APSAM & Argument-principle based stability assessm- \\ 
				& ment method.   \\ 
			IDTA & intersection of the determinant trajectory and \\
			&   axis.  \\
			DT & determinant trajectory.  \\
			FS & frequency sweeping.  \\
			RHP & right half plane. \\
			RTDS & the real-time digital simulation system. \\
		 RSC & 	rotor-side converter. \\
		
		\end{tabular}
\end{table}

\section{Introduction}
\IEEEPARstart{N}{owadays}, the exploitation of wind power has attracted widespread attention, and the doubly-fed induction generator (DFIG) is one of the mainstream wind turbines \cite{intro_wind}. However, several oscillation events of the grid-tied DFIG system have occurred in recent years \cite{intro_sso1}-\cite{intro_sc}, bringing a significant threat to the power system. Considerable studies on the stability analysis of DFIG systems exist.  Most of these studies, however, are based on the assumption that the DFIG system is a white-box system, i.e., the structure and parameters of the system are known, which is not the case in reality. The DFIG system is a grey-box system since the real DFIG's control strategy is private and unavailable to the public \cite{black box1}. In addition, the DFIG system is a typical MIMO system due to the frequency coupling \cite{NC6}, which makes the stability analysis of grey-box DFIG systems challenging.

 Current small-signal stability analysis methods for the grid-tied DFIG systems are mainly divided into the eigenvalue-based and impedance-/admittance-based methods.

The eigenvalue-based methods calculate the eigenvalues and damping ratio of the target system to assess the stability \cite{eig1}-\cite{eig3}.  However, the target system must be a white-box system when utilizing the methods for the stability analysis. Therefore, the eigenvalue-based approaches do not apply to gray-box DFIG systems.

The impedance-/admittance-based methods have been widely used in the stability analysis of wind turbines in recent years \cite{zukang3_1}-\cite{zukang3_3}. These methods construct the frequency response model of wind turbines and then analyze the grid-tied system's stability with the Nyquist stability criterion \cite{NC1},\cite{NC2} or Bode diagrams \cite{NC3},\cite{NC5}. The frequency response model of the device can be obtained in one of two ways. The first is based on theoretical modeling, and the harmonic linearization technique can be used for system modeling. Nevertheless, it necessitates fully comprehending the system structure, parameters, and operating conditions \cite{zukang4},\cite{zukang5}. The second is based on measurement, and several methods have been researched to develop impedance-/admittance- models using frequency sweeping techniques \cite{zukang6},\cite{zukang7}. The latter method can get the impedance model without complete knowledge of the structure and parameters \cite{eig3}. Therefore, producing the wind turbine’s model for the grey-box DFIG system is optimal. Given that the traditional Nyquist criterion and Bode diagram cannot be applied in the MIMO systems, a generalized Nyquist stability criterion (GNC) was proposed in \cite{NC4_1}. This approach is rigorous in theory and accurate, which has been widely used in the stability analysis of the grid-tied DFIG systems \cite{NC6},\cite{NC7}. However, GNC requires the eigenvalue calculation of the return-ratio matrix, which increases the computational burden. Besides, GNC is a qualitative approach to system stability analysis because the dominant modes cannot be directly obtained.

Several attempts have been made to stability analysis of grey-box MIMO systems in recent studies. The impedance measurements are directly converted to state-space models using the vector fitting algorithm in \cite{zong5}. However, the algorithm only converges when the model order is no less than the underlying order, which is unknown to the black-box devices. The Gershgorin circle theorem  is introduced into the stability analysis of MIMO systems in \cite{intro_weak1}, and the computation of the stability criterion can be reduced \cite{pole3}. However, the method is conservative because it needs to set an appropriate prohibited area. The work in \cite{pole1} assesses the system stability according to the determinant-based GNC. The matrix determinant's phase change determines the small-signal stability of the MIMO system. However, its application to the grey-box system needs further exploration as the absolute phase of the frequency response cannot be directly measured. Moreover, the approach cannot be provided for the quantitative analysis. 

This paper proposes an Argument-principle based stability assessment method (APSAM) to perform the stability study of grey-box MIMO systems. The contributions can be summarized as follows:

 \hangafter 1
\hangindent 1.5em
\noindent
1) The argument principle is introduced and customized for the DFIG systems. The proposed method extends the argument principle from single-input-single-output (SISO) systems to MIMO systems utilizing the determinant of the transfer function matrix. Based on the correspondence between argument and determinant trajectory, a simple stability criterion is proposed for DFIG systems.

 \hangafter 1
\hangindent 1.5em
\noindent
2) The method can be applied to the stability analysis of DFIG grey-box systems. The proposed stability criterion relies on the determinant trajectory characteristics of the whole system, which can be derived from the white-box model of the grid and the black-box model of DFIGs. Among them, the black-box model can be built using the frequency sweeping technique.

 \hangafter 1
\hangindent 1.5em
\noindent
 3) The proposed method is simple and suitable for quantitative analysis. The Intersection of Determinant Trajectory and Axis (IDTA) can directly determine the stability, which is straightforward and computationally efficient. Moreover, the method can provide quantitative stability information by evaluating the critical poles. 

The rest of this paper is organized as follows: In section II presents the development of APSAM. Section III gives the stability analysis scheme for the grey-box DFIG system. The proposed method is verified by the time-domain simulation and hardware-in-loop experiment in section IV. Section V discusses and clarifies the concerns, such as  model selection, estimation error analysis and application potential of the proposed method, and Section VI concludes this paper.

\section{Development of APSAM}
\subsection{Argument Principle for the MIMO System}
\begin{figure}[t]
	\centering
	\includegraphics[width=3.2in]{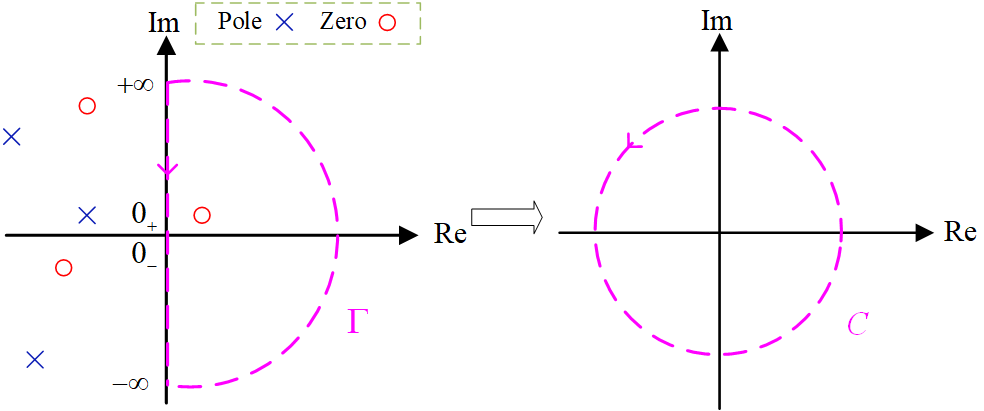}
	\caption{The schematic diagram of the Argument principle.}
	\label{gamma}
\end{figure}

Fig. \ref{gamma} is the schematic diagram of Argument principle. According to \cite{mapping}, the Argument principle can be described as follows:

Denote the complex function as $ w = F(s) $. When the point $ s $ on the complex plane $ S $ goes around once along the closed curve $\Gamma $, the point $ w $ on the complex plane $ W $ plots a continuous closed curve $ C $ accordingly. Define $ \Delta arg F(s) $ as the change in the phase angle of $  F(s) $ on curve $ C $ after one counterclockwise circle of $ s $ along the curve $ \Gamma $, then the difference between the zero number and the pole number of $  F (s) $ in the closed curve $ \Gamma $ equals $ \Delta arg F(s) $ divided by $ 2\pi $.
The principle can be expressed as
\begin{equation}
	\frac{1}{2\pi}\Delta arg F(s)=N(F(s),\Gamma)-P(F(s),\Gamma),
	\label{AP}
\end{equation}
where $ N(F(s),\Gamma) $ and $ P(F(s),\Gamma) $  represent the zero number and the pole number of function $ F(s) $ in the curve $ \Gamma $, respectively. 

According to the matrix theory $ {\cite{Matrix Theory}} $, an $ n $-order MIMO function matrix $ \mathbf{F}(s) $ can be equivalently converted to a diagonal matrix
\begin{equation*}
	\mathbf{F}^{'}(s)=diag\{\lambda_{1}(s), \lambda_{2}(s), \cdots, \lambda_{n}(s)\}.
\end{equation*}
This demonstrates that the zero-pole distribution of a MIMO system whose transfer function is $ \mathbf{F} (s) $ matches that of $ n $ SISO systems with the transfer function $ \lambda_{i}(s) $. In other words, the matrix $ \mathbf{F} (s) $ characteristics depend on its eigenvalues. Based on the Argument principle in (\ref{AP}), for the $ i $-th eigenvalue function $ \lambda_{i}(s) $ of $ \mathbf{F} (s) $,
\begin{equation}
	\frac{1}{2\pi}\Delta arg \lambda_{i}(s)= N(\lambda_{i}(s),\Gamma)-	 P(\lambda_{i}(s),\Gamma).
	\label{Arg_p2}
\end{equation}

The arguments of all eigenvalue functions can be stacked to reflect the zero-pole distribution of the function matrix. All matrix $ \mathbf{F} (s) $'s eigenvalues jointly define its properties.
\begin{equation}
	\sum_{i=1}^{n} \frac{1}{2\pi}\Delta arg \lambda_{i}(s)=	\sum_{i=1}^{n} N(\lambda_{i}(s),\Gamma)-	\sum_{i=1}^{n} P(\lambda_{i}(s),\Gamma).
		\label{Arg_p3}
\end{equation}
Let $\lambda_{i}(s)=r_{i}(s)e^{j\theta_{i}(s)} $, then
\[	\sum_{i=1}^{n}  arg \lambda_{i}(s)=arg\prod_{i=1}^{n} \lambda_{i}(s). \] Considering that multiplying transfer functions do not change the zero-pole distribution of the function, Eq. (\ref{Arg_p3}) can be transformed into
\begin{equation}
	\frac{1}{2\pi}\Delta arg (\prod_{i=1}^{n} \lambda_{i}(s))= N(\prod_{i=1}^{n}\lambda_{i} (s),\Gamma)-	 P (\prod_{i=1}^{n} \lambda_{i}(s),\Gamma).
\end{equation}

Noting that the matrix determinant equals the product of all eigenvalues, the Argument principle to a function matrix can be written as
\begin{equation}
	\frac{1}{2\pi}\Delta arg D(s)= N (D(s),\Gamma)-	 P (D(s),\Gamma),
	\label{Arg_p}
\end{equation}
where $ D(s) $ is the determinant of matrix $ \mathbf{F}(s) $.

Eq. (\ref{Arg_p}) illustrates that the argument of a scalar function $ D(s) $ can reflect the zero-pole distribution of $ \mathbf{F}(s) $. 
\subsection{Customized Argument Principle for the DFIG System}
For a MIMO system, the relationship between its return difference matrix $ \mathbf{F}(s) $ and its closed-loop gain matrix $ \mathbf{T}_{close}(s) $ is
\begin{equation}
	\mathbf{F}(s)={\mathbf{T}_{close}(s)}^{ - 1}.
	\label{Return}
\end{equation} 
According to stability theory \cite{control}, the target MIMO system is unstable when $ \mathbf{T}_{close}(s) $ has right half-plane (RHP) poles or $ \mathbf{F}(s) $ has RHP zeros. Therefore, the system stability can be judged by determining whether $ \mathbf{F}(s) $ has the RHP zero or not. Considering that the distribution of $ \mathbf{F}(s) $'s zeros  coincides with its determinant $ D (s) $ and based on the Argument principle elaborated in (\ref{Arg_p}), the number of zeros in the curve $ \Gamma $ can be expressed by
\begin{equation}
	N (D(s),\Gamma)=\frac{1}{2\pi}\Delta arg D(s)+ P (D(s),\Gamma).
	\label{Arg_z}
\end{equation}

\begin{figure}[t]
	\centering
	\includegraphics[width=2.8in]{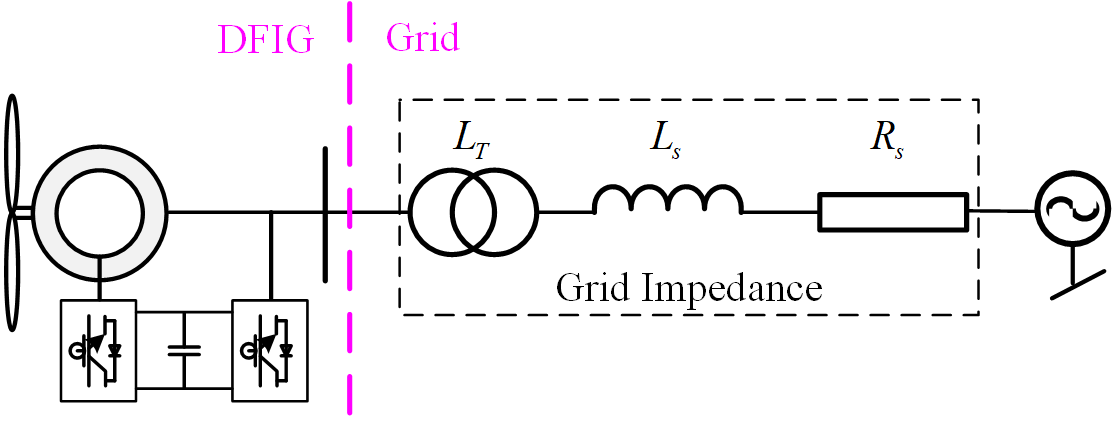}
	\caption{Diagram of DFIG connected to AC grid.}
	\label{DFIG_grid}
\end{figure}

For a typical DFIG system depicted in Fig. \ref{DFIG_grid}, the system can be divided into two subsystems, DFIG and the connected power grid \cite{NC6}\cite{DFIG_adm}. Without losing generality, let
\begin{equation}
	D(s)=det(\mathbf{I}+\mathbf{Z}_{\text {grid }}(s) \mathbf{Y}_{\text{DFIG}}(s)),
\end{equation}
where $ \mathbf{Y}_{\text{DFIG}}(s) $ and $  \mathbf{Z}_{\text {grid }}(s) $ are the MIMO models of DFIG and power grid, respectively; $ \mathbf{I} $ is an identity matrix.

 The system's return-ratio matrix $ \mathbf{G}(s) $ can be expressed as
\begin{equation}
	\mathbf{G}(s) =\mathbf{Z}_{\text {grid }}(s)\mathbf{Y}_{\text{DFIG}}(s) = \left[ {\begin{array}{*{20}{c}}
			{{G_{11}}(s)}&{{G_{12}}(s)}\\
			{{G_{21}}(s)}&{{G_{22}}(s)}
	\end{array}} \right].
\end{equation} 
Take  $ G_{11} (s) $ as an example, and it can be described in the following form
\begin{equation}
	G_{11}(s) = \frac{{\prod\limits_{i = 1}^m {(s - {z_i})} }}{{\prod\limits_{j = 1}^n {(s - {p_j})} }},
\end{equation}
where $ z_i $ and $ p_j $ represent the zero and pole of the function, respectively; $ m $ and $ n $ represent the numerator's and denominator's order, respectively. For a causal system, $ n \ge m $, and
\begin{equation}
D(s) = (1 + {G_{11}}(s))(1 + {G_{22}}(s)) - {G_{12}}(s){G_{21}}(s),
\end{equation}
thus the numerator and denominator are of equal order for $ D(s) $.Therefore, the function $ D(s) $ does not have infinite poles or zeros, and the curve $ \Gamma $ can enclose the zeros of the whole RHP.

In addition, it is believed that $  P ( D(s),\Gamma)=0 $ for the following reasons: 

1) There are no RHP poles in $ \mathbf{Z}_{\text {grid }}(s) $ because the grid contains only passive devices;

 2) No RHP poles exist in $ \mathbf{Y}_{\text{DFIG}}(s) $ since DFIG is self-stable, which is an essential qualification for manufacturers \cite{Self_stable}; 
 
 3) The multiplication or addition of transfer functions does not change the pole distribution.

Under the premise that $ D(s) $ does not contain the RHP poles, Eq. (\ref{Arg_z}) can be transformed into
\begin{equation}
	N (D(s),\Gamma)=\frac{1}{2\pi}\Delta arg D(s)
	\label{Arg_z1}.
\end{equation}
\begin{figure}[t]
	\centering
	\subfigure[$ D(s) $ trajectory.]{
		\includegraphics[width=1.6in]{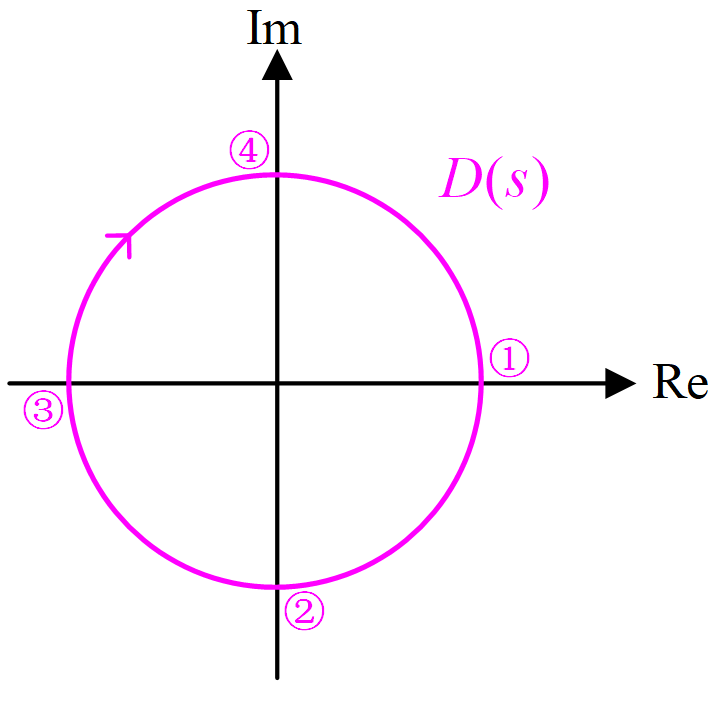} 
		\label{panju1}
	}
	\subfigure[The trajectories of real and imaginary parts.]{
		\includegraphics[width=2.2in]{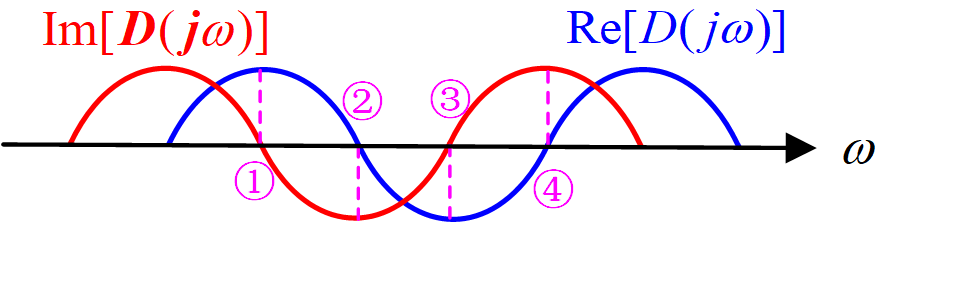} 
		\label{panju2}
	}
	\DeclareGraphicsExtensions.
	\caption{Schematic diagram of the function trajectory.}
\end{figure}
\begin{table}[t]
	\renewcommand\arraystretch{1.1}
	\begin{center}
		\caption{Correspondence between intersection type and value of $ D (s) $}
		\label{tab_1}
		\begin{tabular}{ c c c  c }
			\hline
			Intersection location&Mark & Re$ [D(s)] $ & Im$ [D(s)] $\\
			\hline
			Positive half of real axis	&\ding{172} & $ >0 $  &$  =0 $\\
			\hline
			Negative half of imaginary axis	&\ding{173} & $ =0 $  & $ <0 $\\
			\hline
			Negative half of real axis &\ding{174} & $ <0  $ & $ =0 $\\
			\hline
			Positive half of imaginary axis&\ding{175} & $ =0 $  & $ >0 $\\
			\hline
		\end{tabular}
	\end{center}
\end{table}
For the DFIG system, the zero number of $ D(s) $ in the curve $ \Gamma $ can be reflected by its frequency-domain phase angle variation. It is worth noting that the value range of $ s $ in $ D(s) $ should be set to $ (-j\infty,j\infty) $ to ensure the curve $ \Gamma $ can encompass the entire RHP of the expanded complex plane, and one clockwise rotation of $ D(s) $ around the origin means that $ D(s) $ contains an RHP zero.

As mentioned above, for the DFIG system whose return difference matrix determinant is $ D(s) $, the target system is stable if and only if $ N (D(s),\Gamma) $ in (\ref{Arg_z1}) is equal to 0.

\subsection{Stability assessment based on the D(s) trajectory}
\subsubsection{Qualitative stability assessment}
\label{Trajectory Analysis}

As shown in Fig. \ref{panju1}, when the $D(s)\in \mathbb{C} $ rotates once around the origin clockwise, its trajectory must intersect the coordinate axes of the complex plane in order. The position of the intersection can be judged by the characteristics of the real and imaginary parts of $ D(s) $, such as that Re$ [D(s)]>0 $ and Im$ [D(s)] = 0 $ means $ D(s) $ has an intersection with the positive half of the real axis. The correspondence is shown in TABLE \ref{tab_1}.

The trajectory diagram of Re$ [D(s)] $ and Im$ [D(s)] $ in the frequency domain is presented in Fig. \ref{panju2}, which corresponds to $ D(s) $ trajectory in Fig. \ref{panju1}. Thus, the $ D(s) $ trajectory  is considered to be clockwise around the origin once when different types of intersections appear in order, such as \ding{172}$\rightarrow$\ding{173}$\rightarrow$\ding{174}$\rightarrow$\ding{175}$\rightarrow$\ding{172}. The above analysis shows that combining the characteristics of the real and imaginary parts of $ D(s) $ can introduce $ \Delta arg D(s) $ in (\ref{Arg_z1}).

\begin{figure}[t]
	\centering
	\includegraphics[width=2.0in]{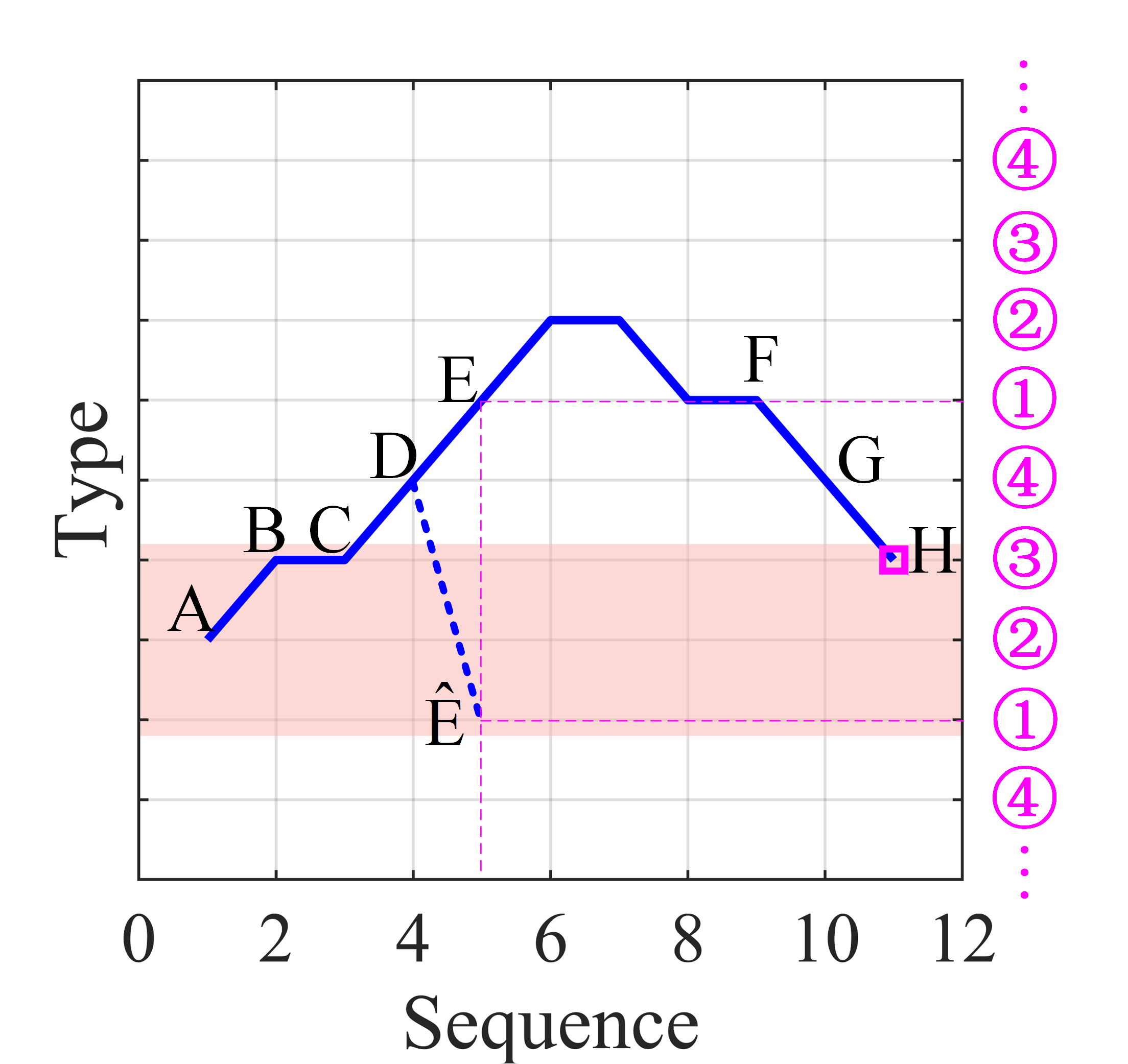}
	\caption{The IDTA diagram. Square: last intersection; Shadow: stable region. }
	\label{IDTA_shiyi1}
\end{figure}

Therefore, the stability of the system whose return difference matrix is $ \mathbf{F}(s) $ can be determined by the Intersection of the Determinant $ D(s) $ Trajectory and the Axis (IDTA). Given that the $ D (s) $ trajectory is a closed curve and that its starting point is unknown, if the first and the last IDTAs are in the same position or adjacent, the trajectory of $ D (s) $ does not encircle the origin and the system is stable; otherwise the opposite. The IDTA curve in Fig. \ref{IDTA_shiyi1} is used to visualize the stability determination results. The plotting principles are as follows:

1) Its horizontal coordinate is the sequence number of IDTA;

2) Its vertical coordinate is the continuous cyclic extension of the intersection type;

3) The vertical coordinates of adjacent intersections are the same or adjacent.

	As shown in Fig. \ref{IDTA_shiyi1}, there are three kinds of curve trends between the adjacent intersections: a) upward increasing by 1: i.e., A $\rightarrow$ B; b) holding flat: i.e., B $\rightarrow$ C; c) downward decreasing by 1: i.e., G $\rightarrow$ H. Especially, if the trends is \ding{175} $\rightarrow$ \ding{172}, the curve should be D $\rightarrow$ E rather than D $\rightarrow$  $ {\rm{\hat E}} $ according to principle 3).

As the first and last IDTAs are in the same position or adjacent under the stable situations, the stable region can be determined as the shadow in Fig. \ref{IDTA_shiyi1} by the first IDTA's location. If the last intersection falls into the stable region, it indicates that the system is stable, and otherwise the opposite.

Theoretically, $ \Delta arg D(s) $ can be obtained based on the phase measurement of $ D(s) $ directly in the whole frequency domain. However, for black-box devices, the measured phase will fall within the interval $ (-\pi,\pi) $, i.e., the phase measurements outside the interval can be wrong, leading to a misjudgment of the grey-box system's stability. In order to correct the measured phase, an additional amount of work is required. In addition, quantitative system stability analysis cannot be performed using phase information alone. In contrast, the stability can be quantitatively assessed by the real and imaginary trajectories of $ D(s) $, which will be described later. Thus, to reflect the phase change utilizing IDTA is chosen rather than applying the measured phase directly.

\subsubsection{Quantitative stability assessment}
Quantitative stability analysis is always a priority in the stability issue.
 The critical pole of the system is defined as the pole closest to the imaginary axis, which can reflect the dominant mode of the system. Here, the critical zero of $ D(s) $ is the critical pole of the system.

Assume that the critical zero of $ D(s) $ is
\begin{equation}
	{z_o} = {\sigma _o} + j{\omega _o},
\end{equation}
where $ \sigma _o $ is the real part of the zero, reflecting the stability margin of the system; $ \omega _o $ is the imaginary part, revealing the potential resonant frequency of the system. According to the stability principle \cite{sigma}, when $ \sigma _o<0 $, the system behavior gradually converges after a small disturbance; when  $ \sigma _o>0 $, the system behavior is gradually divergent. The relationship between the time constant of system behavior and $  \sigma _o $  is
\begin{equation}
|{\sigma _o}| = \frac{1}{\tau },
\end{equation}
where $ |{\sigma _o}| $ represents the absolute value of $ \sigma _o$ and $ \tau $ is the time constant. If $ |{\sigma _o}| $ is large, the corresponding time constant  $ \tau $ is small, indicating faster convergence/divergence of the state variable, otherwise the opposite.

In the small neighborhood of $ z_o $, $ D (s) $ can be expressed as
\begin{equation}
D(s) = (s - {z_o})(a + jb),
\label{D_de}
\end{equation}
where $ a $ and $ b $ are constants, $ s=j\omega $. In the frequency domain, the real and imaginary parts of $ D (j\omega ) $ are
\begin{equation}
\begin{array}{l}
	\vspace{1ex}{\rm{Re}}[D(j\omega )] = ({\omega _o} - \omega )b - {\sigma _o}a,\\
	{\rm{Im}}[D(j\omega )] = (\omega  - {\omega _o})a - {\sigma _o}b.
\end{array}
\label{ex_Re_Im}
\end{equation}
Combine the two equations in (\ref{ex_Re_Im}) to obtain
\begin{equation}
\begin{array}{l}
	{\sigma _o} = \frac{{ - 1}}{{{a^2} + {b^2}}}({\rm{Re}}[D(j\omega )]a + {\rm{Im}}[D(j\omega )]b),\vspace{1.5ex}\\
	{\omega _o} = \omega  + \frac{{{\rm{Re}}[D(j\omega )]b + {\rm{Im}}[D(j\omega )]a}}{{{a^2} + {b^2}}}.
\end{array}
\end{equation}
The critical pole can be calculated at any frequency point in its neighborhood. Without losing generality, the frequency at the imaginary part zero crossing of $ D(j\omega ) $ is used  for computation convenience.
After a simple algebraic calculation, the real and imaginary parts of $ z_o $ can be respectively computed as
 \begin{equation}
\begin{array}{l}
	{\sigma _o} = \frac{{ - {\rm{Re}}[D(j\omega )]a}}{{{a^2} + {b^2}}},\vspace{1.2ex}\\
	{\omega _o} = \omega  - \frac{{{\sigma _o}b}}{a}.
\end{array}
	\label{damping}
\end{equation}
Note that in the neighborhood of $ z_o $, Eq. (\ref{slope}) holds
\begin{equation}
a + jb = \frac{{D(s + j\Delta \omega ) - D(s)}}{{j\Delta \omega }}.
\label{slope}
\end{equation}
Thus, $ a $ and $ b $ can be derived from the slopes of $ D (j\omega) $'s real and imaginary parts.
 \begin{equation}
\begin{array}{l}
	a = \frac{{{\rm{Im}}[D(s + j\Delta \omega )] - {\rm{Im}}[D(s)]}}{{\Delta \omega }},\vspace{1.5ex}\\
	b =  - \frac{{{\rm{Re}}[D(s + j\Delta \omega )] - {\rm{Re}}[D(s)]}}{{\Delta \omega }}.
\end{array}
\label{cal_a_b}
\end{equation}
Therefore, the critical zero of $ D (s) $ can be calculated according to (\ref{damping}).

Only the critical zero of $  D(s) $ can be calculated. If there are no critical zeros of $ D(s) $, i.e., all zeros are far from the imaginary axis, there is no need to quantify the zeros. Because the absence of the critical zero for $ D(s) $ corresponds to two situations: 1) the system is stable and has a large stability margin; 2) the system is extremely unstable and cannot be controlled to pull the system RHP pole back to the left half-plane. In such situations, only the stability of the target system needs to be analyzed qualitatively.

There may be multiple frequencies where $ {\mathop{\rm Im}\nolimits} [D(j\omega)] $ crosses the coordinate axis. However, only one frequency corresponds to the critical state, provided that the critical zero exists for $ D(s) $. When the system is in the critical stable state, $ {\sigma _0} \approx 0 $ can be obtained. According to (\ref{ex_Re_Im}), the 2-norm $ \left\| {(	{\mathop{\rm Re}\nolimits} [D(j\omega )],	{\mathop{\rm Im}\nolimits} [D(j\omega )])} \right\| $ is extremely small in the neighborhood of $ z_o $. Therefore, $ D(s)$'s critical zero is at the frequency corresponding to $ min \left\| {(	{\mathop{\rm Re}\nolimits} [D(j\omega )],	{\mathop{\rm Im}\nolimits} [D(j\omega )])} \right\|  $ \cite{critical zero}. 

\subsection{Steps of APSAM}
\label{pro_APSAM}
For the MIMO system with the return difference matrix $ \mathbf{F}(s) $ (whose determinant is $ D(s) $), the stability analysis steps of the Argument principle based stability assessment method (APSAM) are as follows:

\emph{Step 1}: Plot the IDTA curve based on the trajectories of Re$ [D (j\omega)] $ and Im$ [D (j\omega)] $ in the frequency domain, respectively.

\emph{Step 2}:  Determine the stable region and evaluate the system stability based on the IDTA diagram. If the last intersection falls in the stability region, the system is stable; otherwise, the opposite.

\emph{Step 3}: Calculate the critical zero of $ D (s) $ (the system's critical pole) according to (\ref{damping}) at the frequency  corresponding to $  min \left\| {(	{\mathop{\rm Re}\nolimits} [D(j\omega )],	{\mathop{\rm Im}\nolimits} [D(j\omega )])} \right\|  $. The dominant mode of the system can then be assessed.

\section{Customized Stability Analysis Method for the Grey-box DFIG System}

\subsection{Grey-box Model of the DFIG System}

In the DFIG system, the harmonic components at the frequency $f_{p}$ and $f_{p}-2f_{1}$ always coexist  due to frequency coupling \cite{NC6}. The admittance model of DFIG can be defined by
\begin{equation}
	\mathbf{Y}_{\mathrm{DFIG}}(s)=\left[\begin{array}{ll}
		Y_{11}(s) & Y_{12}(s) \\
		Y_{21}(s) & Y_{22}(s)
	\end{array}\right].
	\label{ex_DFIG}
\end{equation} 
 The specific derivation of the model will not be given here and can be found in Ref. \cite{NC6}.

Obtaining the MIMO model of a black-box DFIG directly by modeling is impossible. However, the frequency-coupled model $ \mathbf{Y}_{\text{DFIG}} $ can be obtained by the frequency sweeping (FS) technique. The specific steps elaborated in Ref. \cite{zong5} are summarized as follows:
	
	1) Provide a small-disturbance voltage signal with variable frequency $ f_p $ at the point of common coupling; 
	
	2) Measure the voltage signals and current signals with the frequency of $ f_p $ and $ f_p-2f_1 $ by FFT; 
	
	3) Calculate the frequency-coupled model based on the measured signals.

The DFIG model is a second-order matrix, so two linear-independent measurements are required. And the calculation method is \begin{small}
	\begin{equation}
		\mathbf{Y}_{\text{DFIG}}(j\omega) = \left[ {\begin{array}{*{20}{c}}
				{{i_{p1,1}}}&{{i_{p1,2}}}\\
				{ {i_{p2,1}}}&{ {i_{p2,2}}}
		\end{array}} \right]{\left[ {\begin{array}{*{20}{c}}
					{ {u_{p1,1}}}&{ {u_{p1,2}}}\\
					{ {u_{p2,1}}}&{ {u_{p2,2}}}
			\end{array}} \right]^{ - 1}}
		\label{TF_obtain},
	\end{equation}
\end{small}
where $ i_{p1,1} $, $ i_{p2,1} $, $ u_{p1,1} $, $ u_{p2,1} $ and $ i_{p1,2} $, $ i_{p2,2} $, $ u_{p1,2} $, $u_{p2,2} $ are the first and second data sets processed by FFT, respectively. It is worth noting that the characteristics of the FS technique determine that the obtained DFIG model is discrete. 
\begin{figure}[t]
	\centering
	\includegraphics[width=2.6in]{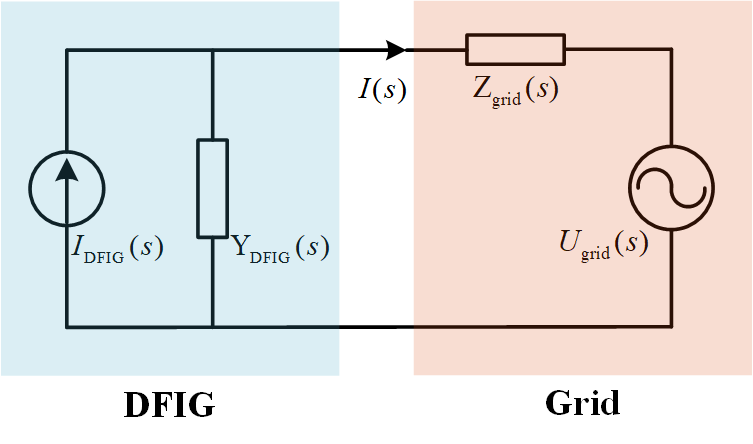}
	\caption{Equivalent simplified model of the grid-tied DFIG system. }
	\label{simplified gridDFIG}
\end{figure}
\begin{figure}[t]
	\centering
	\includegraphics[width=1.6in]{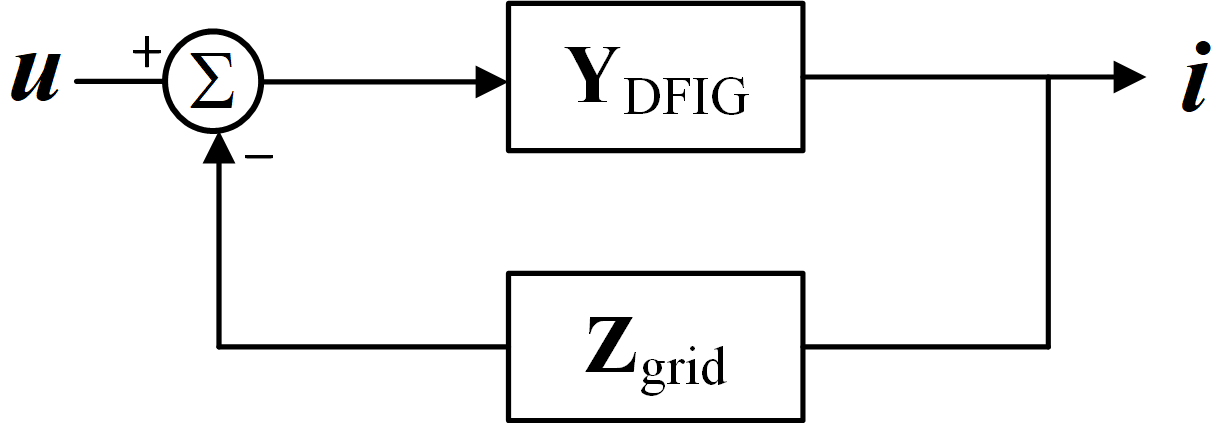}
	\caption{Connection Relationship of the Grid-tied DFIG System. }
	\label{feedback}
\end{figure}

The grid impedance has no frequency coupling effect, and its matrix model shown in (\ref{ex_grid}) has only diagonal elements.
\begin{equation}
	\mathbf{Z}_{\text {grid }}(s)
	=\left[\begin{array}{ll}
		R_{s}+s (L_{s}+L_{T}) & 0 \\
		0 &  R_{s}+s_{2}(L_{s}+L_{T})
	\end{array}\right] ,
	\label{ex_grid}
\end{equation}
where $ R_{s} $, $ L_{s} $, and $ L_{T} $ are the grid resistance, inductance, and equivalent transformer inductance, respectively;  $s_2=s-j2\omega_1$, representing the coupling frequency \cite{NC6}.

The equivalent simplified model of the DFIG grid-connected system is shown in Fig. \ref{simplified gridDFIG}, where the DFIG is equivalent to a Norton circuit with the current source $ {\mathbf{I}_{{\rm{DFIG}}}}(s) $ in parallel with the admittance $ {\mathbf{Y}_{{\rm{DFIG}}}}(s) $, while the grid is equivalent to a Davinan circuit with the voltage source $ {\mathbf{U}_{{\rm{grid}}}}(s) $ in series with the impedance $ {\mathbf{Z}_{{\rm{grid}}}}(s) $ \cite{DFIG_adm}.

Since the DFIG and the power grid are connected in the form of feedback shown in Fig. \ref{feedback}, the system's stability depends on its closed-loop gain $ \mathbf{T}_{close}(s) $.
\begin{equation}
	\mathbf{T}_{close}(s)=	\mathbf{F}(s)^{-1}=\dfrac{1}{\mathbf{I}+\mathbf{Z}_{\text {grid }}(s) \mathbf{Y}_{\text{DFIG}}(s)},
	\label{Sim}
\end{equation} 
where $ \mathbf{I} $ is the 2-order identity matrix.

\begin{figure}[b]
	\centering
	\includegraphics[width=3.2in]{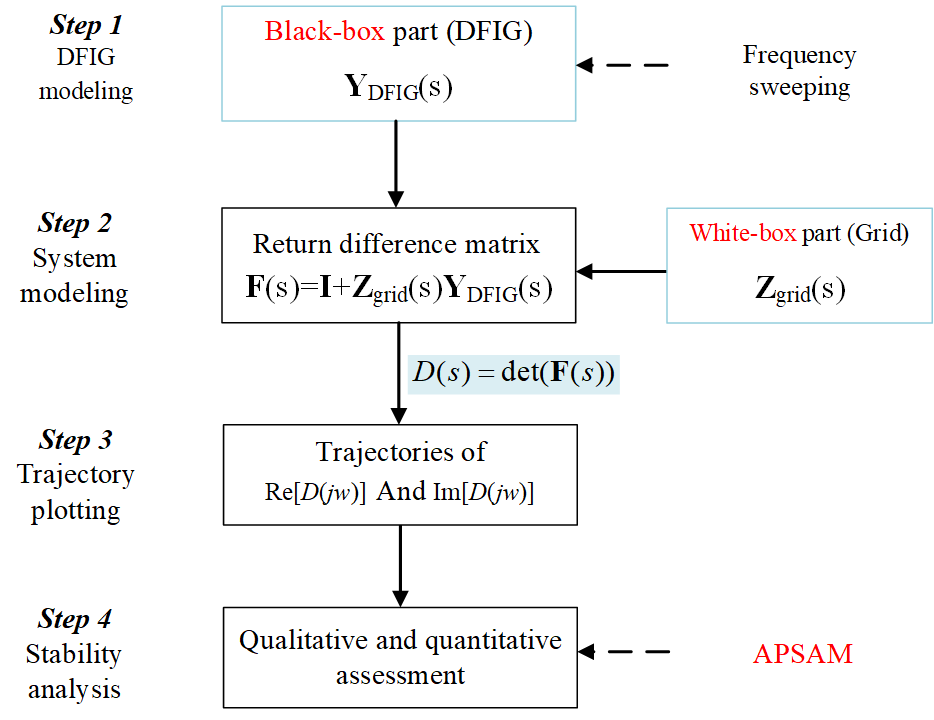}
	\caption{The stability assessment approach for the grey-box DFIG system with APSAM. }
	\label{flowchart_APSAM}
\end{figure}

\subsection{Tips for Applying the Method}

\subsubsection{Frequency range setting}
In the frequency domain, choosing a frequency range of $ (-\infty, \infty) $ for FS is not implementable. Based on the results of numerous studies of DFIG integrated into AC grid systems, it is reasonable to set the FS range at (-1000, 1000) Hz \cite{NC6},\cite{zukang3_3}. In addition, to reduce the workload when carrying out FS, the sweeping frequencies at (0, 100) Hz should be as dense as possible, while the density of the frequencies in the other ranges can be moderately reduced.
\subsubsection{Over-zero point judging}
The transfer functions are discrete when the target system is a grey box. When analyzing the characteristics of $ D(s) $ trajectory, $ D(s) $ is considered to have an intersection with the coordinate axis when the discrete points take values that satisfy the following conditions:
\begin{equation}
{\mathop{\rm Re}\nolimits} [D(j{\omega _i})]{\mathop{\rm Re}\nolimits} [D(j{\omega _{i + 1}})] \le 0,
\label{OPJ1}
\end{equation} 
 or 
 \begin{equation}
 {\mathop{\rm Im}\nolimits} [D(j{\omega _i})]{\mathop{\rm Im}\nolimits} [D(j{\omega _{i + 1}})] \le 0,
 \label{OPJ2}
\end{equation} 
where $ \omega _i $ and $ \omega _{i+1} $  represent the angular frequency of the $  i $-th and $  (i+1) $-th sweeping point, respectively.
\subsubsection{Frequency interval processing}
The obtained model based on FS is discrete and has frequency intervals. When calculating the critical pole, the method of piecewise linear interpolation is adopted to process the data. Then the parameters in Eq. (\ref{damping}) can be acquired to calculate the system's critical pole.

\subsection{Stability Analysis Scheme for the Grey-box DFIG System}

For the grey-box DFIG system, the stability analysis scheme shown in Fig. \ref{flowchart_APSAM} is as follows:

\emph{Step 1}: Obtain the frequency-coupled model $ \mathbf{Y}_{\text{DFIG}}(s) $ of the black-box DFIG by FS technique;

\emph{Step 2}: Calculate the return  difference matrix $ \mathbf{F}(s) $ of the system combining the white-box model $ \mathbf{Z}_{\text {grid }}(s) $ and the black-box model $ \mathbf{Y}_{\text{DFIG}}(s) $ obtained in \emph{Step 1};

\emph{Step 3}: Calculate the determinant $ D (s) $ of $ \mathbf{F}(s) $ and draw the Re$ [D (s)] $ and Im$ [D (s)] $ trajectories in the frequency domain;

\emph{Step 4}: Analyze the stability of the grey-box DFIG system according to APSAM proposed in Subsection \ref{pro_APSAM}.

 \section{Method Verification}
\begin{table}[b]
	\renewcommand\arraystretch{1.1}
	\begin{center}
		\caption{Parameters of the Grid-tied DFIG system in simulation}
		\label{tab_2}
		\begin{tabular}{ c  c  c }
			\hline
			Parameter & Symbol & Value\\
			\hline
			Rated Voltage & $ U_{1} $ & 0.69kV\\
			
			Rated Power & $ S $ & 1.5MVA \\ 
			
			Rated Frequency & $f_{1}$ & 50Hz \\
			
			Stator Resistance & $ R_{s} $ & 0.0090838p.u.
			\\
			Rotor Resistance & $ R_{r} $ &  0.009015p.u. \\
			Stator Leakage Inductance & $ L_{s} $ &  0.18167p.u \\
			Rotor Leakage Inductance & $ L_{r} $ &  0.143969p.u. 
			\\
			Mutual Inductance  & $ L_{m} $ &  5.8959p.u. 
			\\
			GSC Filter Inductance & $ L_{c} $ &  0.0005H 
			\\
			Rotor Speed & $ \omega_{m} $ & 0.8p.u. 
			\\
			Equivalent Reactance of Transformer & $ Z_{T} $ & 0.65p.u. 
			\\
			Equivalent Reactance of Grid & $ Z_\text{grid} $ & 0.001p.u. 
			\\
			PLL Proportional Gain & $ K_{p p} $ & 50 
			\\
			PLL Integral Gain & $ K_{p i} $ & 100 \\
			RSC Proportional Gain
			& $ K_{r p} $ & 0.2
			\\
			RSC Integral Gain
			& $ K_{r i} $ & 10 
			\\
			RSC Decoupling Gain & $ K_{r d} $ & -0.34 
			\\
			GSC Proportional Gain & $ K_{c p} $ & 0.5 
			\\
			GSC Integral Gain & $ K_{c i} $ & 5 
			\\
			GSC Decoupling Gain & $ K_{c d} $ & -0.157 
			\\
			\hline 
			Line Resistance & $ R_{line} $ & $ 7.98e^{-3} \Omega  $
			\\
			Line Inductance & $ L_{line} $ &  $ 2.68e^{-4}$ H    
			\\
			\hline
		\end{tabular}
	\end{center}
\end{table}
The APSAM's accuracy is first verified by time-domain simulation, and then the effectiveness of the proposed method for grey-box DFIG systems is experimentally confirmed.

\subsection{Simulation Verification}

\label{sim_case}

\begin{table}[b]
	\renewcommand\arraystretch{1.2}
	\begin{center}
		\caption{Parameters of the critical pole for the simulation}
		\label{tab_omega_damp_simulation}
		\begin{tabular}{ c | c  c  c c c }
			\hline
			$ K_{rp} $	 & $ f $ & Re[$ D(s) $] & $ a $ & $ b $ &$ z_o $ of $ D(s) $\\
			\hline
			0.15& 55.5Hz &0.0102871 &0.11617&-0.12930&-0.15790+348.89i\\
			0.1& 55.1Hz &-0.0820905 &0.19828&-0.15026&0.26387+345.99i\\
			\hline
		\end{tabular}
	\end{center}
\end{table}

 A grid-tied DFIG system shown in Fig. \ref{DFIG_grid} is built on the PSCAD/EMTDC platform, and its main parameters are displayed in TABLE~\ref{tab_2}. The return difference matrix $ \mathbf{F}(s) $ of the target system and its determinant $ D(s) $ can be calculated by combining the derived DFIG model in (\ref{ex_DFIG}) with the grid model in (\ref{ex_grid}).

\subsubsection{Analysis by the APSAM}

When a DFIG is connected to the power grid, its rotor-side converter (RSC) current inner loop parameters have a significant impact on system stability \cite{intro_weak1}. Therefore, the DFIG system under different RSC current inner loop parameter $ K_{rp} $ is analyzed  by the APSAM here. $ K_{rp} $ is set to 0.15 and 0.1 to study the system's stability, respectively.

\begin{figure}[t]
  	\centering
  	\subfigure[DT curve and the corresponding intersection types. Red: imaginary curve; Blue: Real curve. Circle number: Intersection type.]{
  		\includegraphics[width=0.9\linewidth]{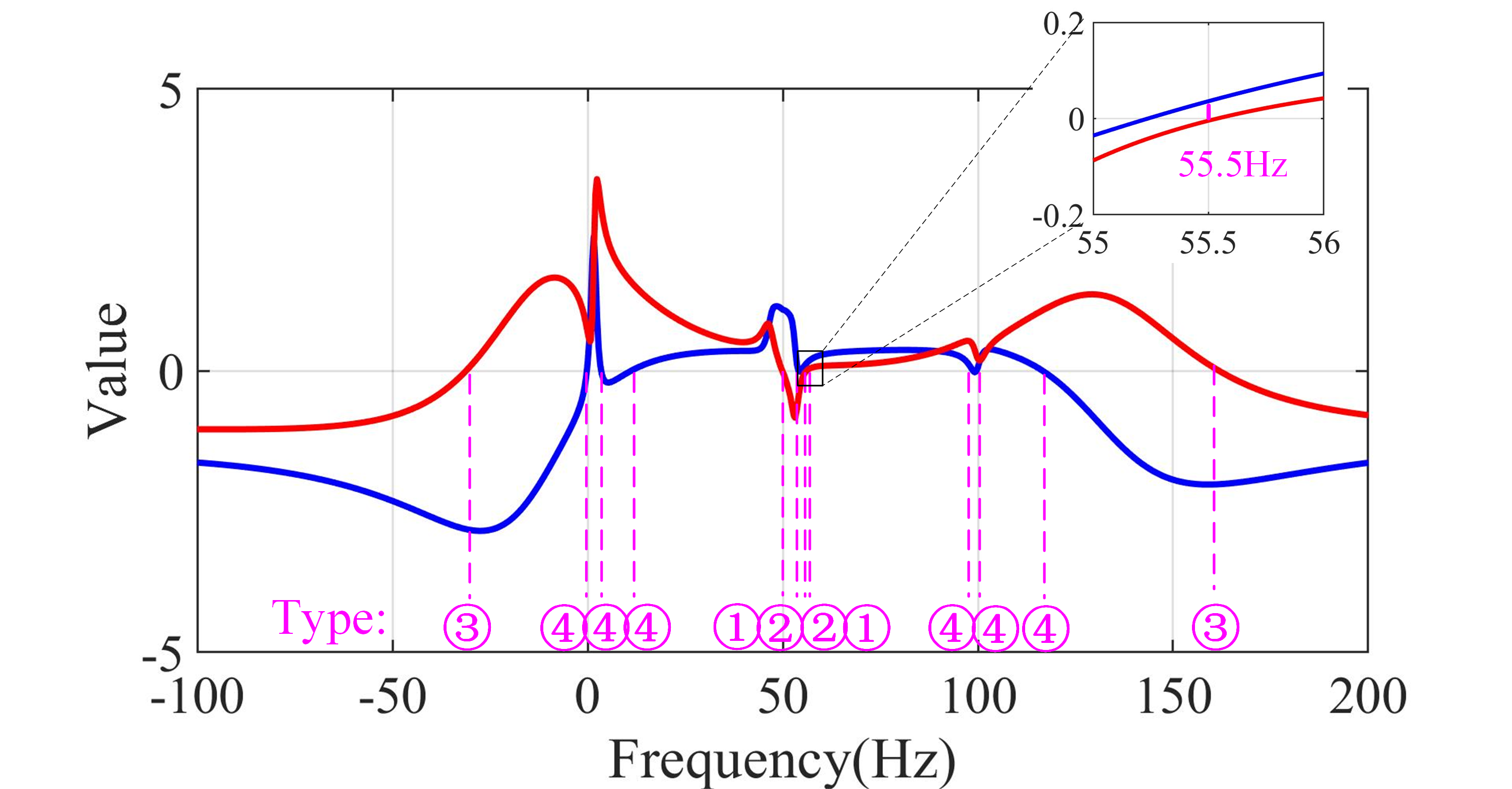} 
  		\label{APSAM_and_zero_type _0.15}
  		
  	}
  	\subfigure[The trajectory characteristics of $ D(s) $. Left: Trajectory diagram; Right: IDTA diagram. Square: last intersection; Shadow: stable region.]{
  		\includegraphics[width=0.83\linewidth]{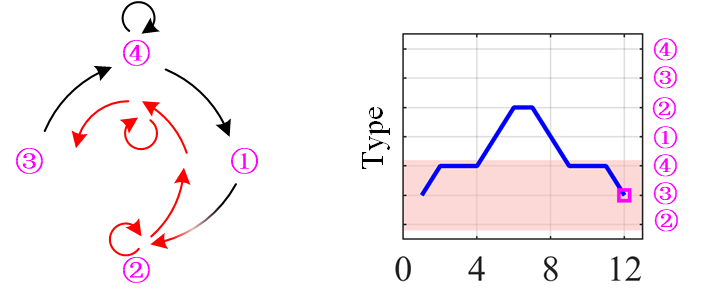} 
  		\label{zero_type_shiyi0.15}
  	}
  	\DeclareGraphicsExtensions.
  	\caption{Stability analysis based on APSAM when $ K_{rp}=0.15 $.}
  	\label{APSAM_curve0.15}
  \end{figure}

 Fig. \ref{APSAM_curve0.15}  shows the determinant trajectory (DT) curve when $ K_{rp}=0.15 $, and the corresponding relationship between the DT curve and different intersections is shown in Fig. \ref{APSAM_and_zero_type _0.15}. The left panel of Fig. \ref{zero_type_shiyi0.15} is used to visually represent the trajectory characteristics of $ D(s) $, and the right panel shows the IDTA diagram, which is used to determine the system stability automatically by the computer. 
 
 Based on the IDTA diagram in  Fig. \ref{zero_type_shiyi0.15}, it can be seen that when $ K_{rp} $ is set to 0.15,  $ D(s) $ has a total of 12 IDTAs, and the last intersection falls into the stable region. That means $  D(s) $ does not contain RHP zeros, and the target system is stable.

\begin{figure}[t]
  	\centering
  	\subfigure[DT curve and the corresponding intersection types. Red: imaginary curve; Blue: Real curve. Circle number: Intersection type.]{
  		\includegraphics[width=0.9\linewidth]{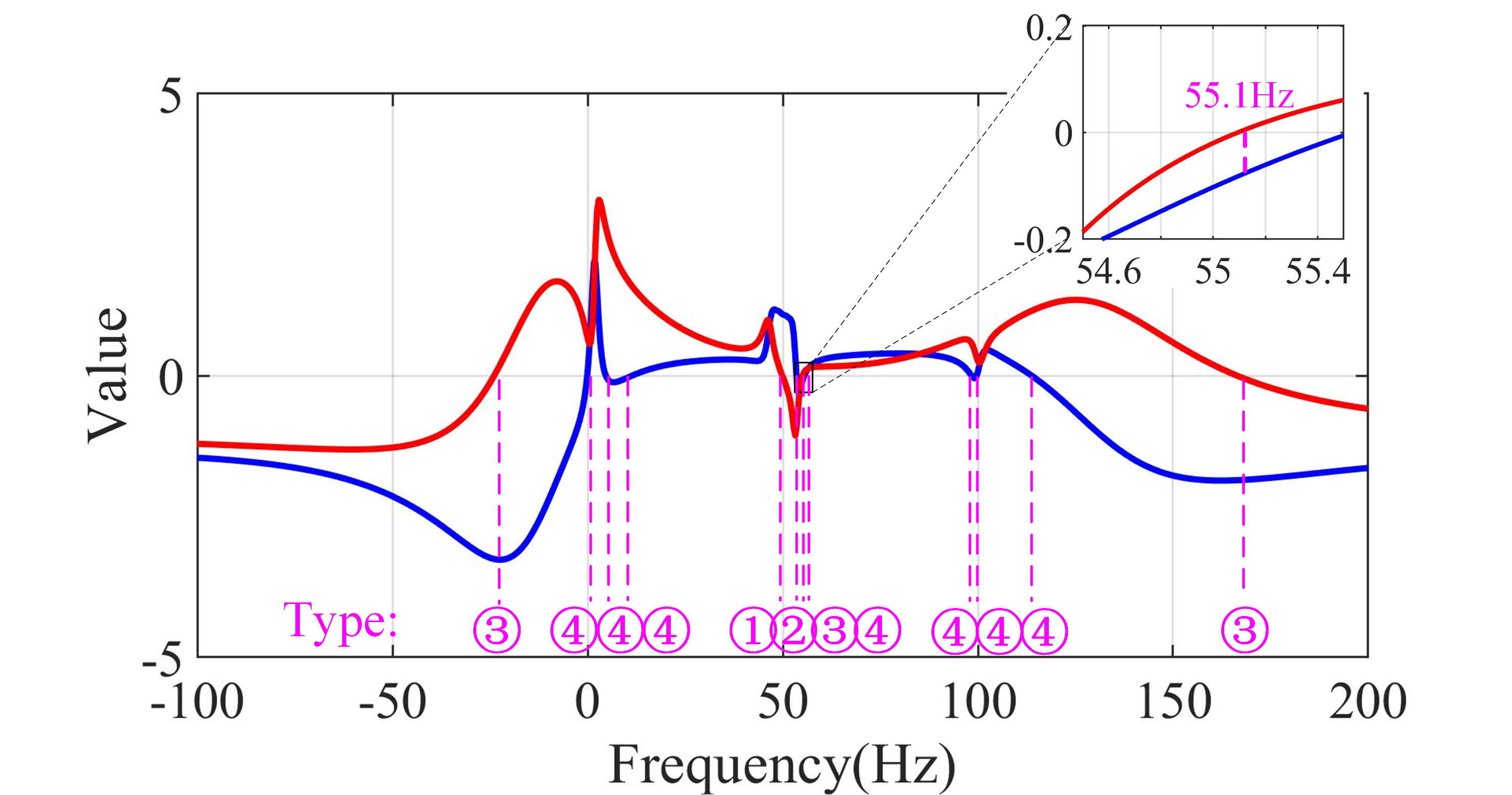} 
  		\label{APSAM_and_zero_type _0.1}
  		
  	}
  	\subfigure[The trajectory characteristics of $ D(s) $. Left: Trajectory diagram; Right: IDTA diagram. Square: last intersection; Shadow: stable region.]{
  		\includegraphics[width=0.8\linewidth]{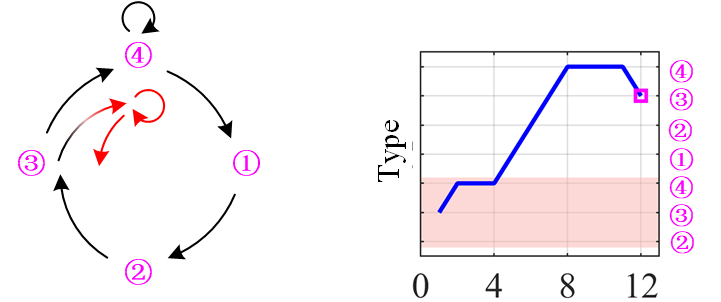} 
  		\label{zero_type_shiyi0.1}
  	}
  	\DeclareGraphicsExtensions.
  	\caption{Stability analysis based on APSAM when $ K_{rp}=0.1 $.}
  	\label{APSAM_curve0.1}
  \end{figure}

  Fig. \ref{APSAM_curve0.1}  shows the DT curve when $ K_{rp}=0.1 $. In the IDTA diagram shown in Fig. \ref{zero_type_shiyi0.1}, the last intersection falls outside the stable region. It indicates that $ D(s) $ has an RHP zero, and the system is unstable.
  \begin{figure}[t]
	\centering
	\includegraphics[width=3.2in]{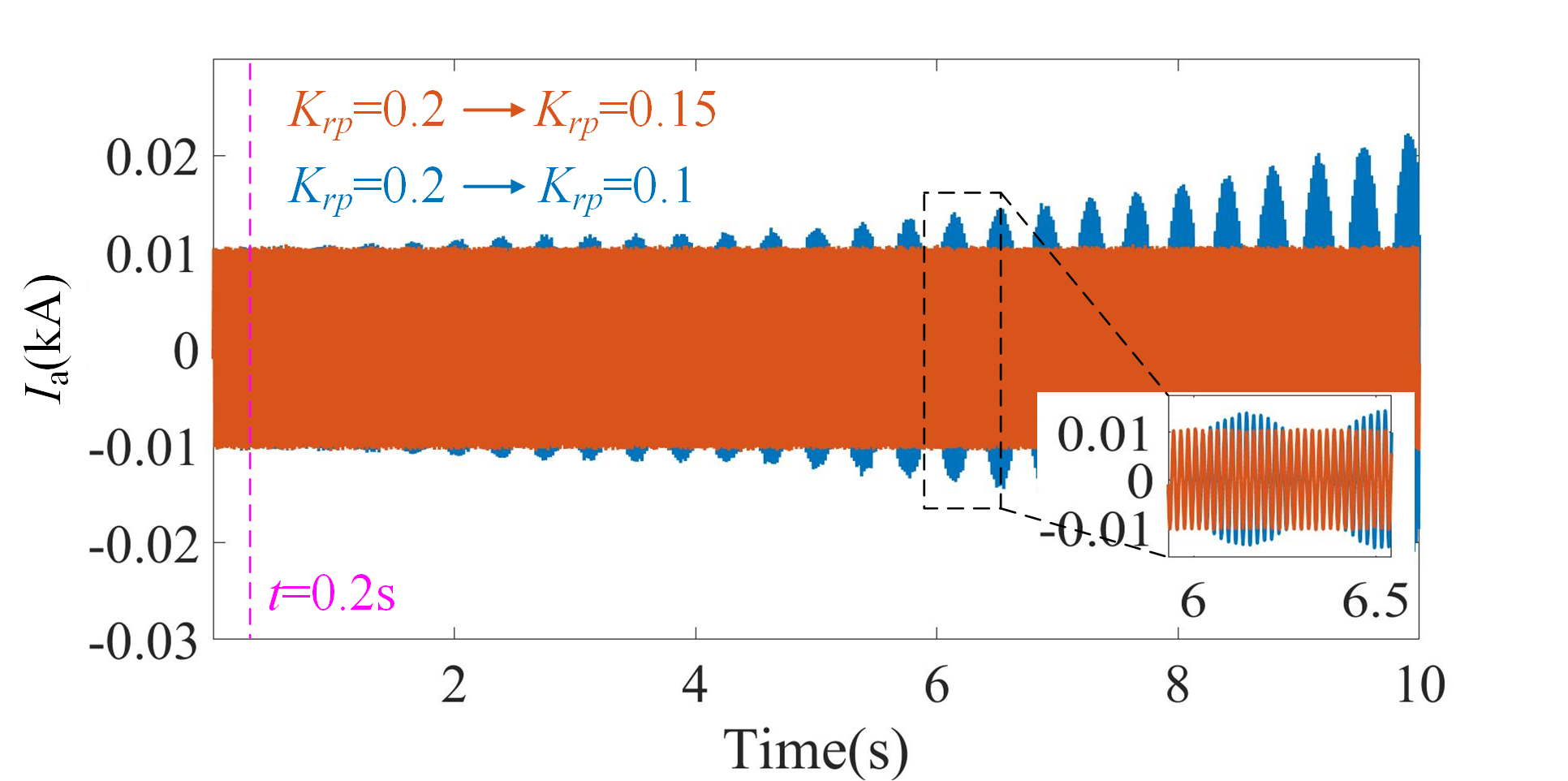}
	\caption{The simulated current waveforms when the $ K_{rp} $ is changed. }
	\label{Current_wave_2_case}
\end{figure}

\begin{figure}[t]
	\centering
	\subfigure[$K_{rp}=0.1$]{
		\includegraphics[width=0.8\linewidth]{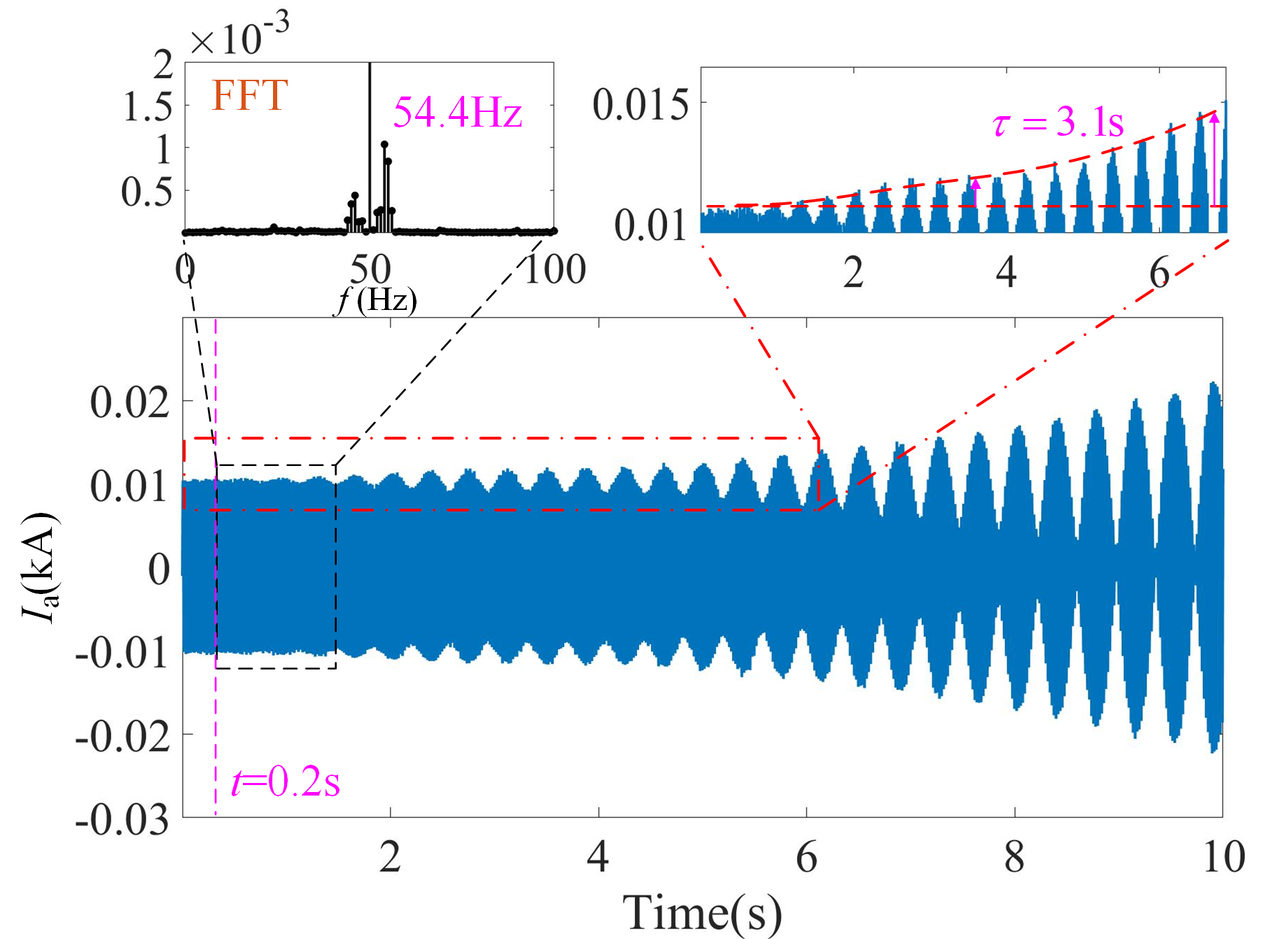} 
		\label{simulation1}
		
	}
	\subfigure[$K_{rp}=0.15$. Apply voltage disturbance in ($ t_1 $, $ t_2 $).]{
		\includegraphics[width=0.8\linewidth]{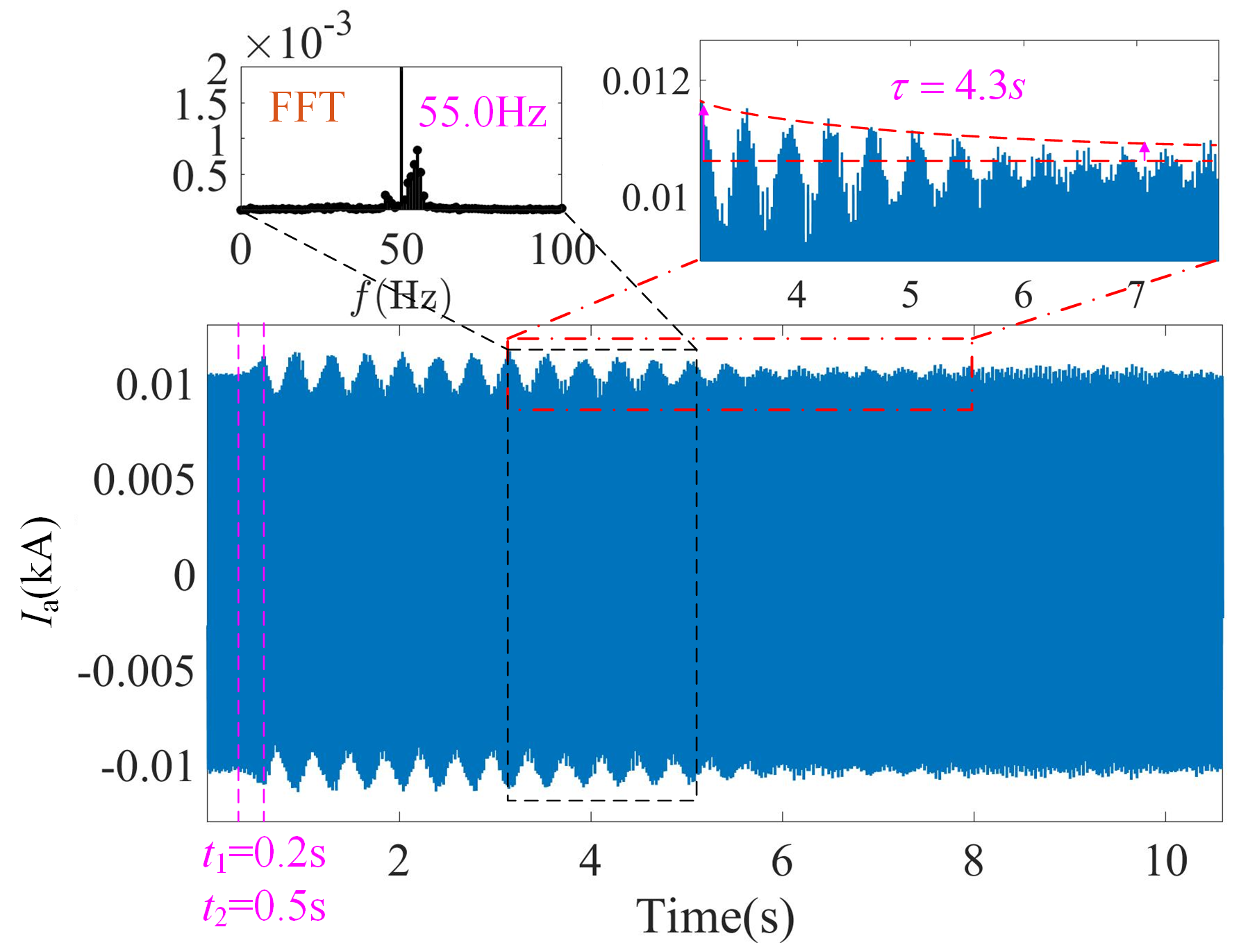} 
		\label{simulation2}
	}
	\DeclareGraphicsExtensions.
	\caption{The A-phase current waveforms.}
	\label{simulation}
\end{figure}
 \begin{figure}[t]
 	\centering
 	\subfigure[]{
 		\includegraphics[width=0.8\linewidth]{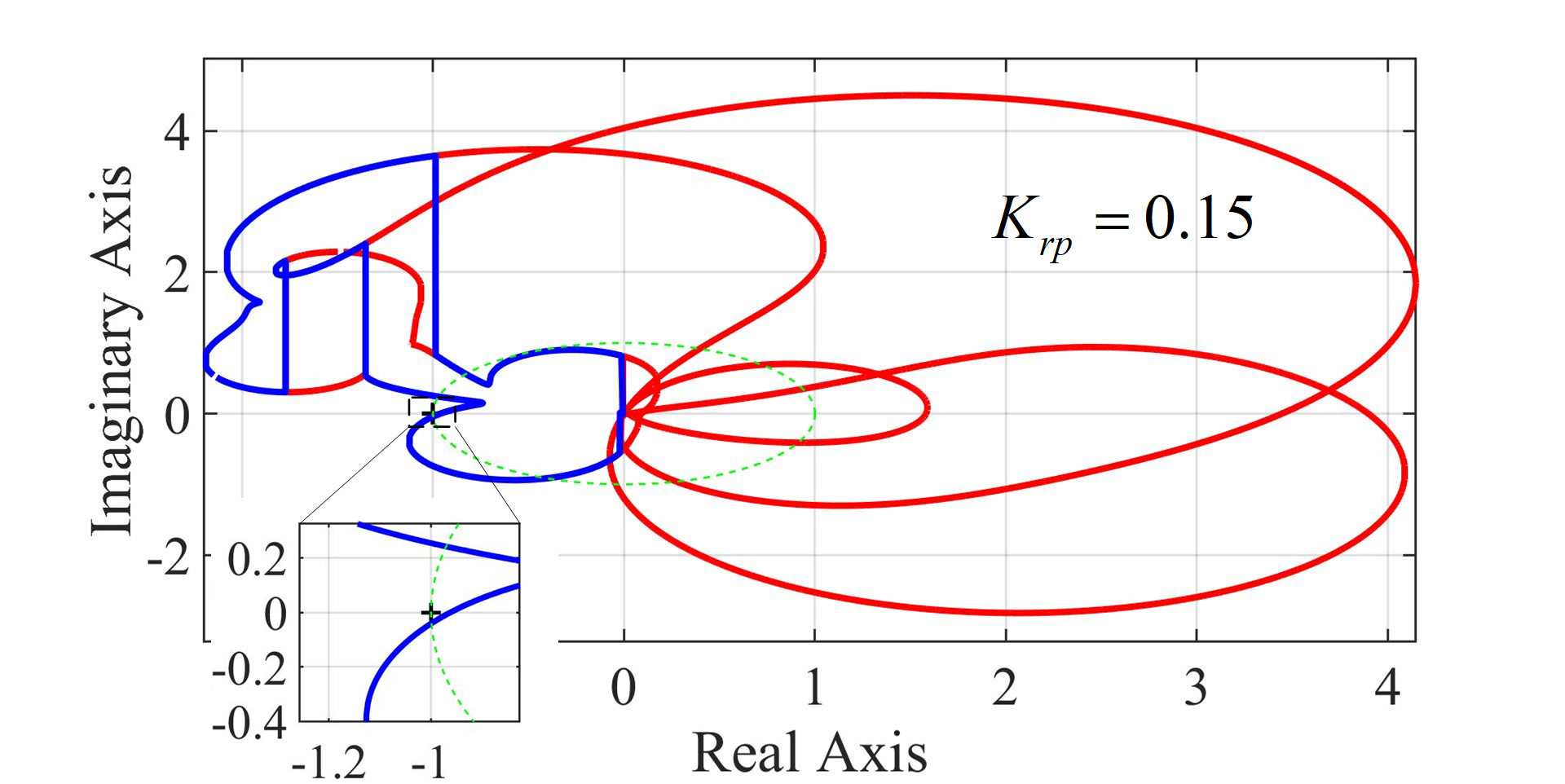} 
 		\label{GNC0.15}
 		
 	}
 	\subfigure[]{
 		\includegraphics[width=0.8\linewidth]{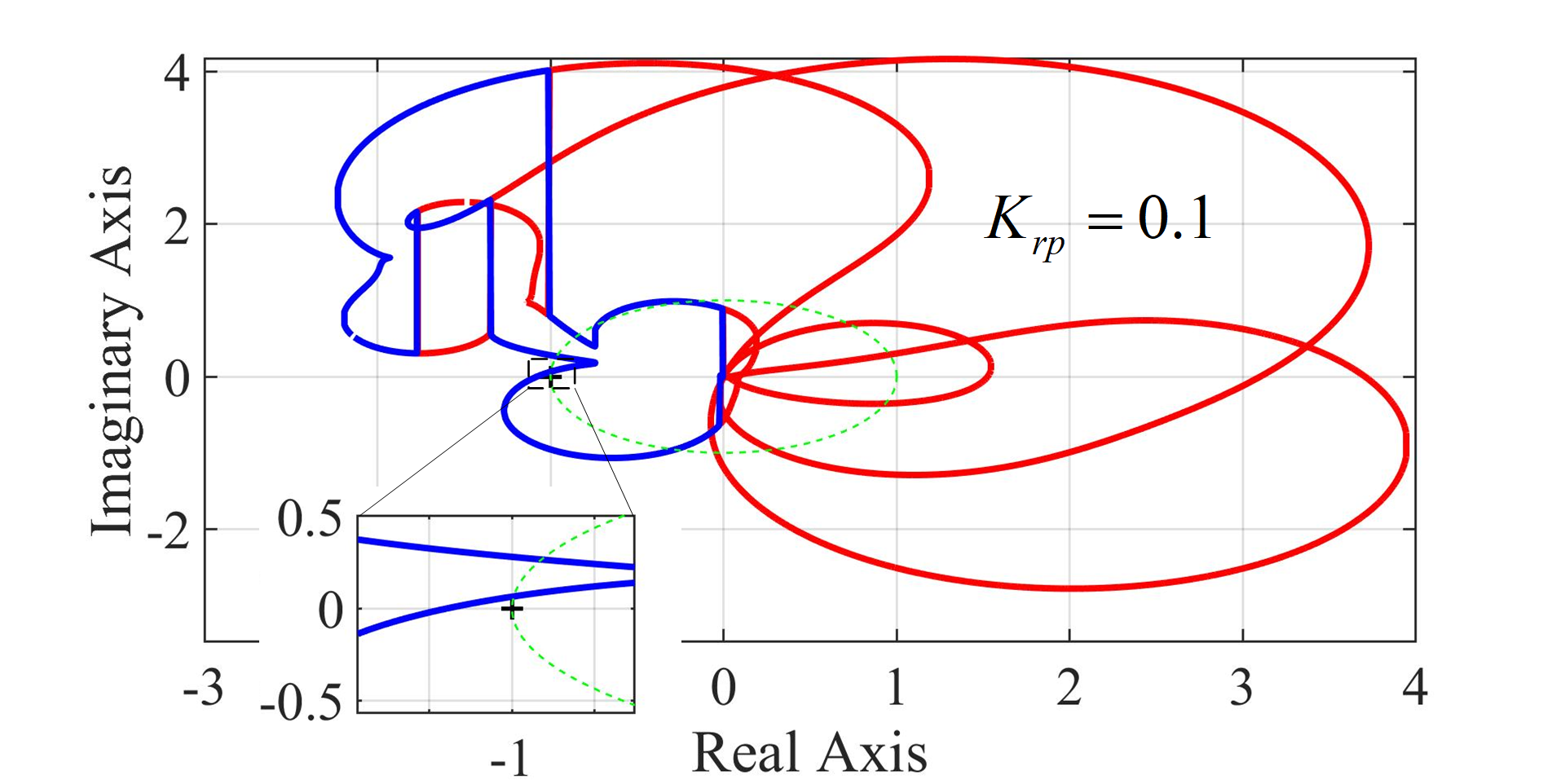} 
 		\label{GNC0.1}
 	}
 	\DeclareGraphicsExtensions.
 	\caption{GNC curves for the DFIG system under different $ K_{rp} $. Blue: $ \lambda_{1} $; Red: $ \lambda_{2} $.}
 	\label{GNC}
 \end{figure}

 \begin{figure}[t]
 	\centering
 	\includegraphics[width=2.5in]{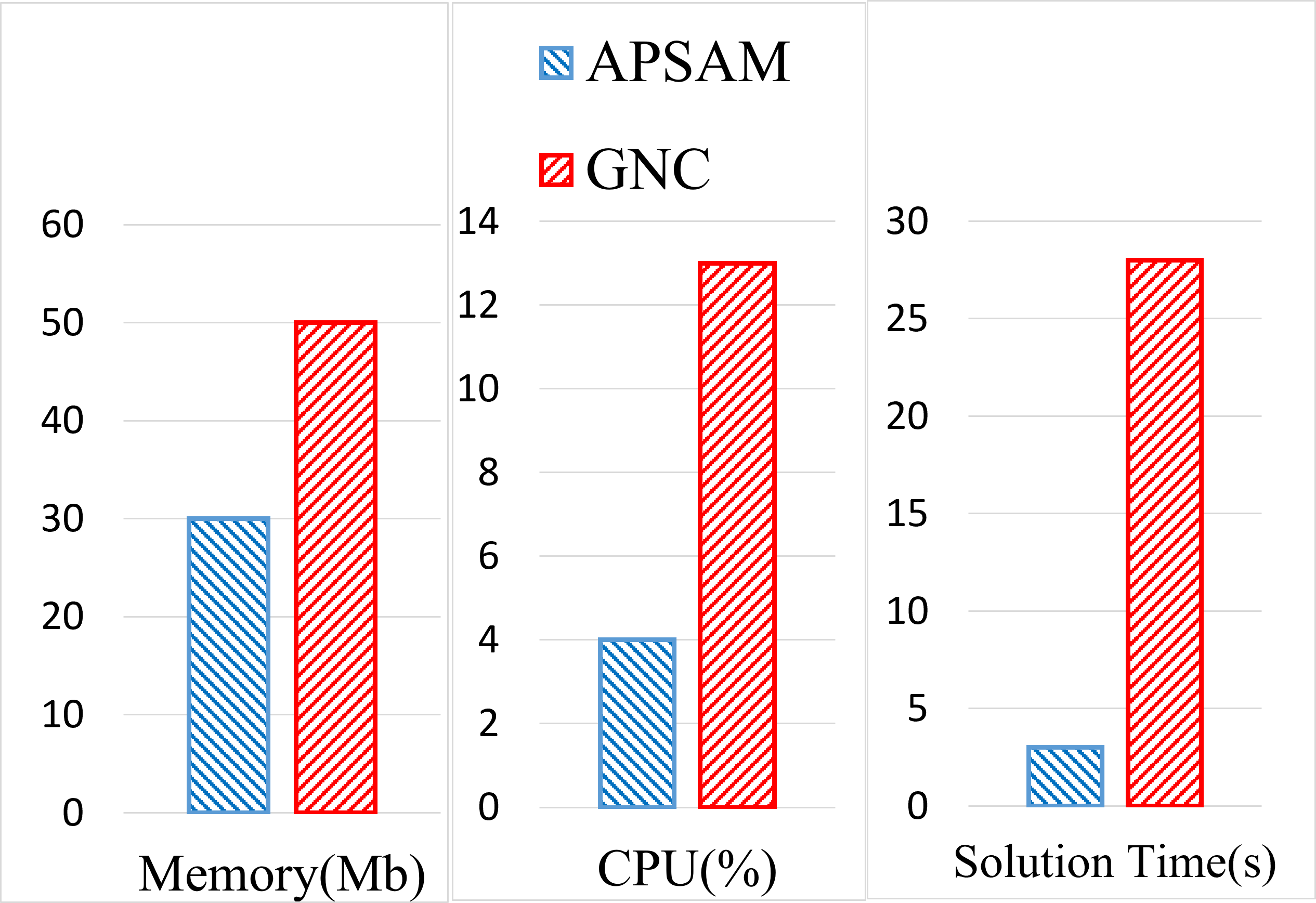}
 	\caption{Comparison of the computational burden of APSAM and GNC. }
 	\label{APSAM_GNC_compare}
 \end{figure}

From Fig. \ref{APSAM_and_zero_type _0.15} and Fig. \ref{APSAM_and_zero_type _0.1}, it can be found that the system's critical pole corresponds to the frequency of about 55.5Hz and 55.1Hz, respectively. The critical pole can be calculated by Eq. (\ref{damping}), and TABLE \ref{tab_omega_damp_simulation} displays the relevant parameters and the calculated critical zero of $ D(s) $ under different $ K_{rp} $. Accordingly, the system is stable when $ K_{rp}=0.15 $, and $ K_{rp}=0.1 $ causes the system to oscillate at about 55.06Hz (345.99/2$ \pi $).

\subsubsection{Time-domain simulation}
The stability evaluation results are validated by time-domain simulations. Under the parameter conditions shown in TABLE \ref{tab_2}, the system is in stable operation.  Only the A-phase current is used for the demonstration here.

Initially, the system is operating steadily. At 0.2s, $ K_{rp} $ is adjusted to 0.15 and 0.1, respectively. Fig. \ref{Current_wave_2_case} shows the current waveforms under different parameter conditions. It can be seen that when $ K_{rp} $ is changed to 0.15, the current waveform remains stable. When $ K_{rp} $ is changed to 0.1, the current waveform has a significant oscillation. This suggests that the qualitative determination of system stability using APSAM is correct.

Fig. \ref{simulation1} shows the current waveform when $ K_{rp}=0.1 $. An FFT analysis of the waveform during 0.2s-1.2s shows a 54.4 Hz frequency component in addition to the fundamental frequency component. The analysis of the overall waveform amplitude variation demonstrates that the time constant of the oscillation component is about 3.1 s.
In Fig. \ref{simulation2}, $ K_{rp} $ is set to 0.15, and a voltage disturbance is superimposed to the system from 0.2s to 0.5s. It can be found that the current waveform gradually converges after the disturbance. According to FFT analysis results and waveform convergence characteristics, the system's natural resonance frequency is about 55.0Hz, and the time constant is 4.3s.

 The above simulations show that the critical poles of the system are about $ 0.32+j343.43 $ and $ -0.23+j345.58 $ respectively when $ Krp=0.1 $ and $ Krp=0.15 $, while the calculated critical poles using APSAM are $ 0.26+j345.99 $ and $ -0.16+j348.89 $, respectively. Considering the errors in data measurement and processing, the calculated critical pole can reflect the dominant mode of the system, and the results of APSAM for the quantitative analysis of the system stability are accurate.
 
\begin{figure}[t]
	\centering
	\includegraphics[width=3in]{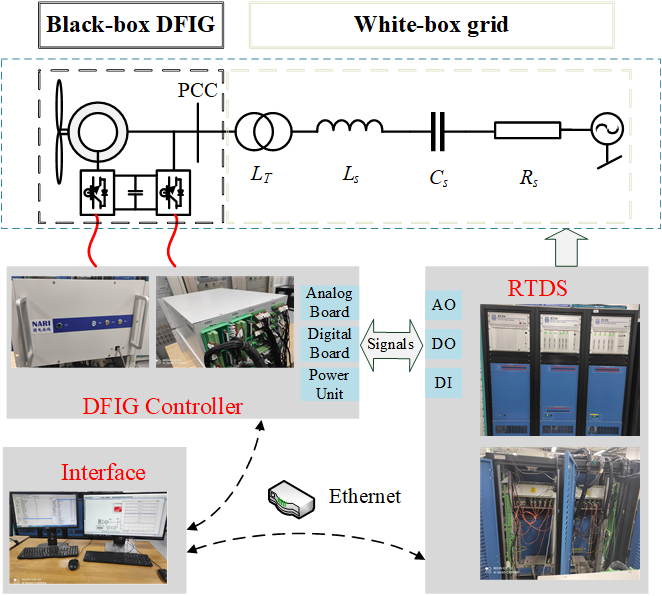}
	\caption{Grey-box DFIG system based
		on the hardware-in-loop simulator.}
	\label{RTDS}
\end{figure}

\subsubsection{Compared with GNC}
The generalized Nyquist criterion (GNC) is used to examine the stability of the DFIG system to confirm the accuracy of the stability evaluation results by APSAM. The GNC curves are displayed in Fig. \ref{GNC}. When $ K_{rp}=0.15 $, the curves do not surround the point $ (-1, j0) $, meaning the system is stable. When $ K_{rp}=0.1 $, the trajectory of $ \lambda_{1} $ encompasses the point $ (-1, j0) $, indicating that the target system contains an RHP pole. In this situation, the system's characteristic function $ D(s) $ contains an RHP zero. The analysis conclusions by GNC are consistent with the assessment results of the APSAM. Moreover, compared with the conventional GNC, the APSAM can avoid counting the turn number of GNC curves around the point $ (-1, j0) $, which is tedious when the system is complex \cite{pole1}.

\begin{table}[h]
	\renewcommand\arraystretch{1.2}
	\begin{center}
		\caption{Parameters of the Grid-tied DFIG System on RTDS}
		\label{tab_3}
		\begin{tabular}{ c  c  c }
			\hline
			Parameter & Symbol & Value\\
			\hline
			Rated voltage of power grid & $ U_{\text{grid}} $ & 35kV\\
			Rated Voltage of DFIG  & $ U_{\text{DFIG}} $ & 0.69kV\\
			Converter DC voltage & $ U_{dc} $ & 1.08kV \\
			Rated Power & $ S $ & 2.1MVA \\ 
			
			Rated Frequency & $f_{1}$ & 50Hz \\
			
			Stator Resistance & $ R_{s} $ & 0.008086p.u.
			\\
			Rotor Resistance & $ R_{r} $ &  0.006616p.u. \\
			Stator Leakage Inductance & $ L_{s} $ &  0.0978p.u \\
			Rotor Leakage Inductance & $ L_{r} $ &  0.1707p.u. 
			\\	Mutual Inductance & $ L_{m} $ &  4.718p.u. 
			\\
			Rotor Speed & $ \omega_\text{m} $ & 0.8p.u. 
			\\
			
			Equivalent Reactance of Transformer & $ Z_\text{T} $ & 0.065p.u. 
			\\
			\hline
		\end{tabular}
	\end{center}
\end{table}

 Fig. \ref{APSAM_GNC_compare}  compares the memory, CPU usage, and solution time for stability analysis between APSAM and GNC by solving 5000 frequency points. The complexity (memory), CPU usage, and solution time of APSAM are smaller than those of GNC, which is because APSAM avoids the eigenvalue computation of the function matrix involved in GNC \cite{intro_weak1}. This indicates that the proposed APSAM is much less computationally complex than the conventional GNC when analyzing the stability of MIMO systems.

\subsection{Experimental Verification}
 
\subsubsection{Experimental platform}
To confirm the effectiveness of the proposed APSAM for the grey-box DFIG system, a DFIG grid-tied system is built on a hardware-in-loop experimental platform, with its schematic diagram shown in Fig. \ref{RTDS}. Some parameter settings are shown in TABLE \ref{tab_3}.
\begin{figure}[t]
	\centering
	\includegraphics[width=1.0\linewidth]{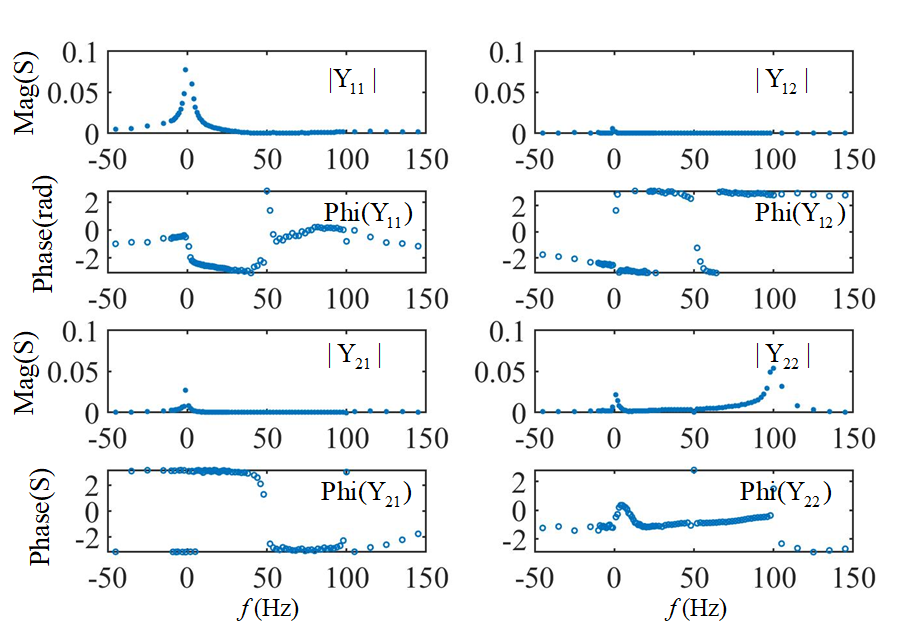}
	\caption{Black-box DFIG frequency-coupled model obtained by FS.}
	\label{Black_DFIG}
\end{figure}

\begin{figure}[t]
	\centering
	\subfigure[DT curves for the grey-box target system in the experiment]{
		\includegraphics[width=0.9\linewidth]{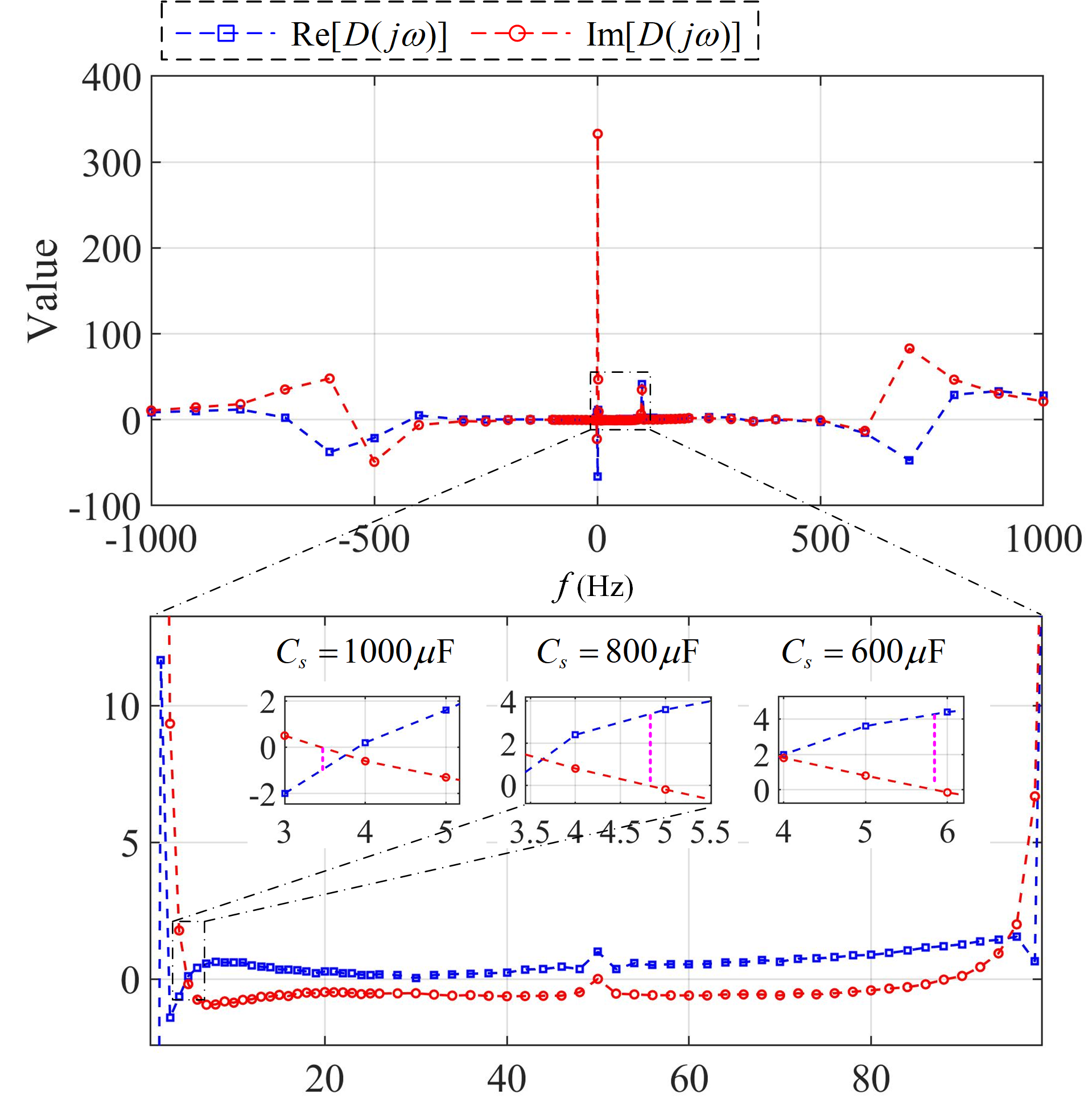}
		\label{APSAM_curve_grey}
	}
	\subfigure[Schematic of the $ D(s) $ trajectories]{
		\includegraphics[width=2.6in]{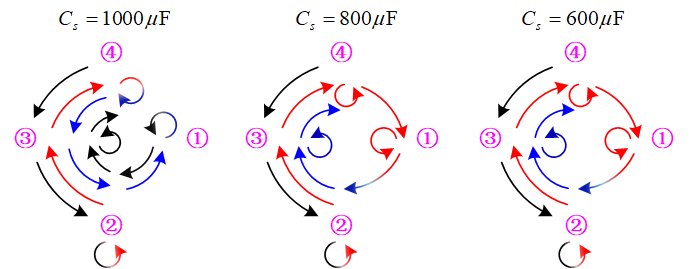} 
	}
	\subfigure[IDTA diagram. Square: last intersection; Shadow: stable region.]{
		\includegraphics[width=2.6in]{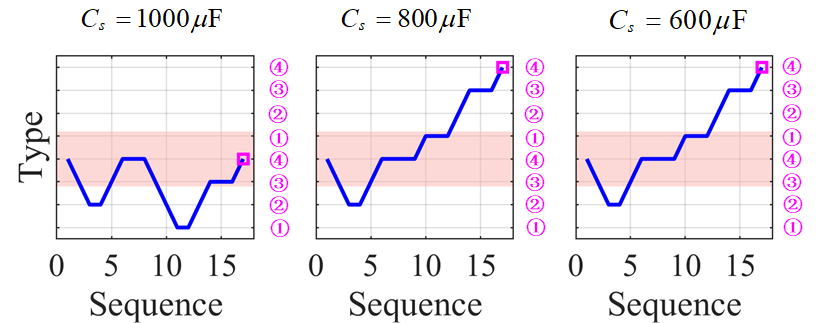} 
		\label{intersection_sys_IFTC}	
	}
	\DeclareGraphicsExtensions.
	\caption{The trajectory characteristics of $ D(s) $ under different $ C_{s} $.}
	\label{intersection_sys}
\end{figure}

The controller of DFIG adopts the actual controller produced by NARI Tech. in China. The control structure and parameters of the DFIG are not known, so the DFIG is a black-box device, and the whole system is a grey-box system. The rest of the target system, including the induction machine, the transmission line, and the transformer, are simulated on the real-time digital simulation system (RTDS). The computer exchanges information with RTDS and DFIG controller through Ethernet, and signals are transmitted through a physical connection between RTDS and the actual controller. This method ensures the reliability of the analysis results to the greatest extent.

\subsubsection{Analysis by the APSAM}
According to (\ref{TF_obtain}), the black-box DFIG frequency-coupled model can be obtained based on the FS technique, shown in Fig. \ref{Black_DFIG}. Only the FS results of (-50, 150) Hz are shown in the figure for clear presentation. Combing the black-box DFIG model with the white-box power grid model (\ref{ex_grid}), the target grey-box system's characteristic function $ D(s) $ can be obtained.

 The stability analysis of the target system is carried out by altering the value of the series compensation capacitor $ C_s $ in the white-box grid. Fig.\ref{intersection_sys} shows the DT characteristics under different $ C_s $. According to the IDTA diagram in Fig. \ref{intersection_sys_IFTC}, when $ C_s $ is 1000$ \mu $F, the last intersection falls into the stable region, indicating that there is no unstable pole and the system is stable. When $ C_s $ is changed to 800$ \mu $F and 600$ \mu $F, the last intersection falls outside the stable region, indicating that the system has an RHP pole and the system will oscillate.

Under the condition that $ min \left\| {(	{\mathop{\rm Re}\nolimits} [D(j\omega )],	{\mathop{\rm Im}\nolimits} [D(j\omega )])} \right\|  $ holds at the system's critical pole, the corresponding frequency is about 5 Hz. The specific DT curves obtained from FS for the grey-box target system are displayed in Fig. \ref{APSAM_curve_grey}. Due to the FS technique's properties, the produced DT curves of the grey-box system are discrete. According to Eq. (\ref{damping}), TABLE \ref{tab_omega_damp} provides the estimated critical pole.

\begin{figure*}[t]
	\centering
	\includegraphics[width=0.985\linewidth]{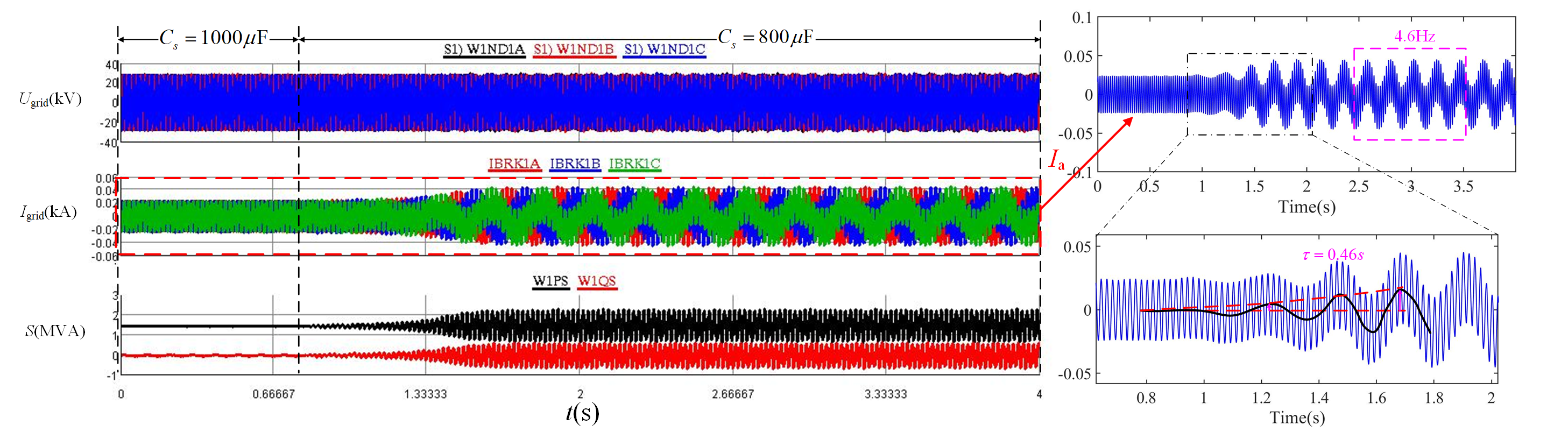}
	\caption{Experimental waveforms of the system when $ C_s $ is changed to 800$ \mu $F.}
	\label{fig:Experimental_800}
\end{figure*}

\begin{figure*}[t]
	\centering
	\includegraphics[width=1.0\linewidth]{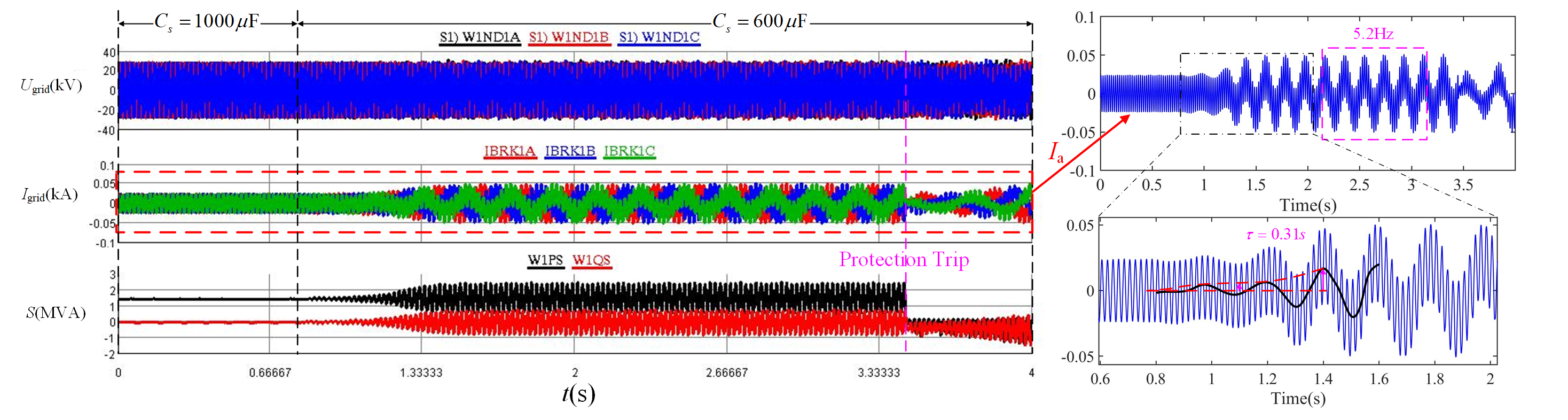}
	\caption{Experimental waveforms of the system when $ C_s $ is changed to 600$ \mu $F.}
	\label{fig:Experimental_600}
\end{figure*}

When $ C_s=1000\mu$F, the real part of the system's critical pole is negative, and the target system  is stable. When $ C_s=800\mu $F and $ C_s=600\mu $F, the real part of the system's critical pole is positive, indicating the system is unstable. The corresponding oscillation frequencies are $ 4.532$ Hz(28.475/2$ \pi $) and $ 5.503$ Hz(34.579/2$ \pi $), respectively. 

\subsubsection{Experimental results}

Experiments are conducted to verify the stability evaluation results by altering $ C_s $ of the system during the stable operation of the grey-box DFIG system. Fig. \ref{fig:Experimental_800} and Fig. \ref{fig:Experimental_600} show the waveforms of the system after changing $ C_s $ from 1000$ \mu $F to 800$ \mu $F and 600$ \mu $F at 0.8s. The waveforms include the grid voltage $ U_{\text{grid}} $, the grid current $ I_{\text{grid}} $, and the delivered power $ S $, respectively.

The system waveforms in Fig.\ref{fig:Experimental_800} exhibit an oscillation of approximately 4.6Hz following the adjustment of $ C_s $, and the real part of the pole corresponding to the time constant should be 2.17(1/0.46). In Fig. \ref{fig:Experimental_600}, the system waveforms exhibit an oscillation of roughly 5.2Hz following the modification of $ C_s $, and the real part of the critical pole should be 3.23(1/0.31).

The oscillation waveforms of the system at $ C_s=600\mu $F dissipate faster, and the oscillation amplitude is more significant than that at $ C_s=800\mu $F, and even triggers the action of the relay protection, which is consistent with the findings of the system damping by the APSAM. Compared with the calculation results in TABLE \ref{tab_omega_damp}, the assessment results for the target system's damping and oscillation frequency are acceptable.
\begin{table}[b]
	\renewcommand\arraystretch{1.2}
	\begin{center}
		\caption{The calculated critical pole of the system for the experiments}
		\label{tab_omega_damp}
		\begin{tabular}{ c | c  c  c c c }
			\hline
			$ C_s $	 & $ f $ & Re[$ D(s) $] & $ a $ & $ b $ &$ z_o $ of $ D(s) $\\
			\hline
			1000$ \mu $F& 3.52Hz &-1.051 &-0.967&1.715&-0.262+22.556i\\
			800$ \mu $F& 4.80Hz &3.327 &-0.985&1.131&1.456+28.475i\\
			600$ \mu $F& 5.84Hz &4.397 &-1.012&0.772&2.747+34.579i\\
			\hline
		\end{tabular}
	\end{center}
\end{table}

Note that the DFIG's frequency coupling in Fig. \ref{Black_DFIG} is weak, and ignoring off-diagonal admittance seems to have no effect on the stability analysis. However, the weak frequency coupling of DFIG is only an example and does not have universality.  Actually, some studies such as Ref. \cite{NC6} and Ref. \cite{MIMO_modified} have expounded that DFIG's off-diagonal admittance has a significant effect on the stability analysis. Besides, Appendix \ref{stability_mis} gives a stability misjudgment case when ignoring off-diagonal admittance. Therefore, it is recommended to take into account the DFIG's off-diagonal admittance in the stability analysis of grey-box DFIG systems.
 
\section{Discussion}

\subsection{DFIG Admittance / Impedance Model}

When using the proposed APSAM to assess the stability of grey-box DFIG systems, the sweeping DFIG's admittance $ \mathbf{Y}_{\text{DFIG}}(s) $ rather than the impedance $ \mathbf{Z}_{\text{DFIG}}(s) $ is adopted. This is because the FS technique's characteristics determine the selection of the admittance. When performing FS on the black-box DFIG, the input signal $ u $ is the voltage signal $ v $, and the output signal $ y $ is the current signal $ i $. Accordingly, the transfer function $ H_{\rm{DFIG}} $ is
 \begin{equation}
{H_{{\rm{DFIG}}}} = \frac{y}{u} = \frac{i}{v} = {{\rm{Y}}_{{\rm{DFIG}}}}.
\end{equation} 
 Thus, the use of the admittance form is more consistent with the FS technique's physical meaning.

Even if the DFIG's impedance model is chosen, stability analysis can still be done using the proposed APSAM. The discussed DFIG is self-stable because this is an essential qualification for manufacturers \cite{Self_stable}, thus the DFIG can be regarded as bounded-input-bounded-output (BIBO) stable \cite{BIBO_stable} on the small disturbance. Combined with  $ {{\bf{Z}}_{{\rm{DFIG}}}}(s) = {{\bf{Y}}_{{\rm{DFIG}}}}{(s)^{ - 1}} $, it indicates that neither $ \mathbf{Y}_{\text{DFIG}}(s) $ nor $ \mathbf{Z}_{\text{DFIG}}(s) $ has any RHP poles or zeros. For the return difference matrix of the DFIG grid-connected system, the following equation holds.
 \begin{equation}
{\bf{I}} + {{\bf{Z}}_{{\rm{grid }}}}(s){{\bf{Y}}_{{\rm{DFIG}}}}(s) = {\bf{C}}(s)[{\bf{I}} + {{\bf{Z}}_{{\rm{DFIG }}}}(s){{\bf{Y}}_{\rm{grid}}}(s)],
\label{im_ad_eq}
 \end{equation}
where $ {\bf{C}}(s)={{\bf{Z}}_{{\rm{grid }}}}(s){{\bf{Y}}_{{\rm{DFIG}}}}(s) $; $ \mathbf{Z}_{\text{grid}}(s) $ has no RHP poles or zeros since the power grid only contains passive components. Eq. (\ref{im_ad_eq}) implies that the RHP pole-zero number of $ [{\bf{I}} + {{\bf{Z}}_{{\rm{grid }}}}(s){{\bf{Y}}_{{\rm{DFIG}}}}(s)] $  and  $ [{\bf{I}} + {{\bf{Z}}_{{\rm{DFIG }}}}(s){{\bf{Y}}_{\rm{grid}}}(s)] $ is equal. And it can be verified that the stability conclusions obtained by adopting the two models, $ \mathbf{Y}_{\text{DFIG}}(s) $ and $ \mathbf{Z}_{\text{DFIG}}(s) $, are consistent given the cases in \ref{sim_case}. Appendix \ref{model_selec} gives the specific verification.

Therefore, adopting the admittance model  $ \mathbf{Y}_{\text{DFIG}}(s) $  or the impedance model $ \mathbf{Z}_{\text{DFIG}}(s) $ for the DFIG system can yield the correct stability result.

\subsection{Estimation Errors}
\emph{1) Error analysis}

Though the proposed method can estimate the critical poles by (\ref{damping}) to realize the stability analysis of the grey-box system, estimation errors are inevitable. The errors are mainly divided into theoretical errors and errors caused by frequency sweeping.

(1) Theoretical errors. Although the critical poles have a decisive influence on the dynamic process of the system, other non-critical poles will also have a small but inevitable impact on the system's dynamic performance, which will bring errors to the critical poles analyzed by the simulations and experiments.

(2) Errors caused by FS:

a) Errors caused by disturbance signals. When performing FS, small-disturbance voltages are superimposed on the system. According to the engineering experience, the voltage amplitude of the disturbance signal is less than 5\% of the operating voltage. However, small disturbances in engineering and its mathematical definition are different. The small superimposed disturbances may affect the accuracy of the measurement model to a certain extent.

b) Errors caused by noise. When FS is carried out, the measurement noises are inevitable. In practical application, the error caused by noise can be alleviated by improving the performance of the FS device.

c) Errors caused by frequency intervals. The frequency intervals of FS determine the accuracy of the admittance model. Smaller frequency intervals can benefit fewer errors at the cost of heavier frequency sweeping workload and computational burden. 

\begin{figure}[t]
	\centering
	\subfigure[$ K_{rp}=0.15$]{
		\includegraphics[width=0.8\linewidth]{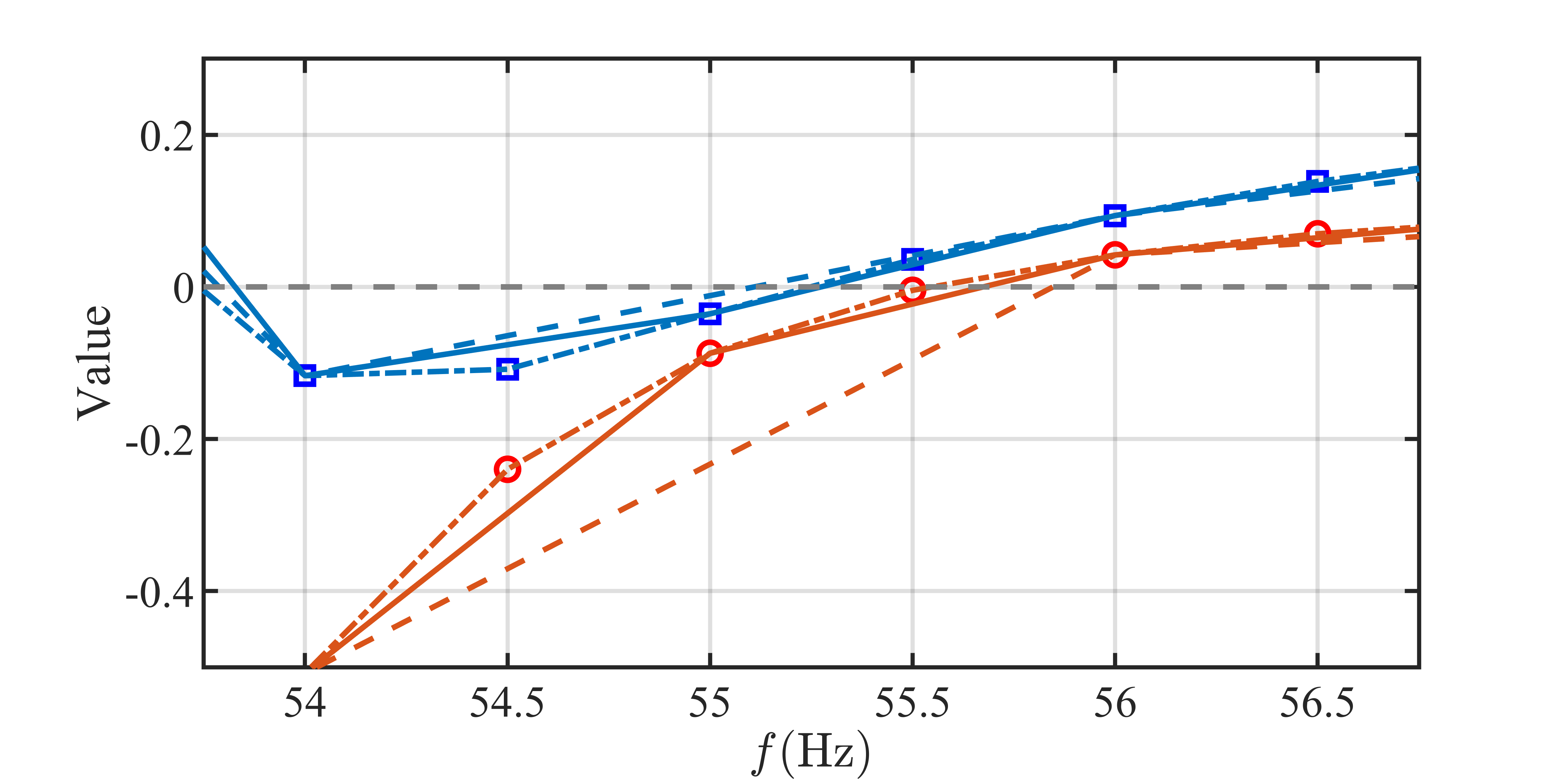}
		\label{f_jiange015}
	}
	\subfigure[$ K_{rp}=0.1$]{
		\includegraphics[width=0.8\linewidth]{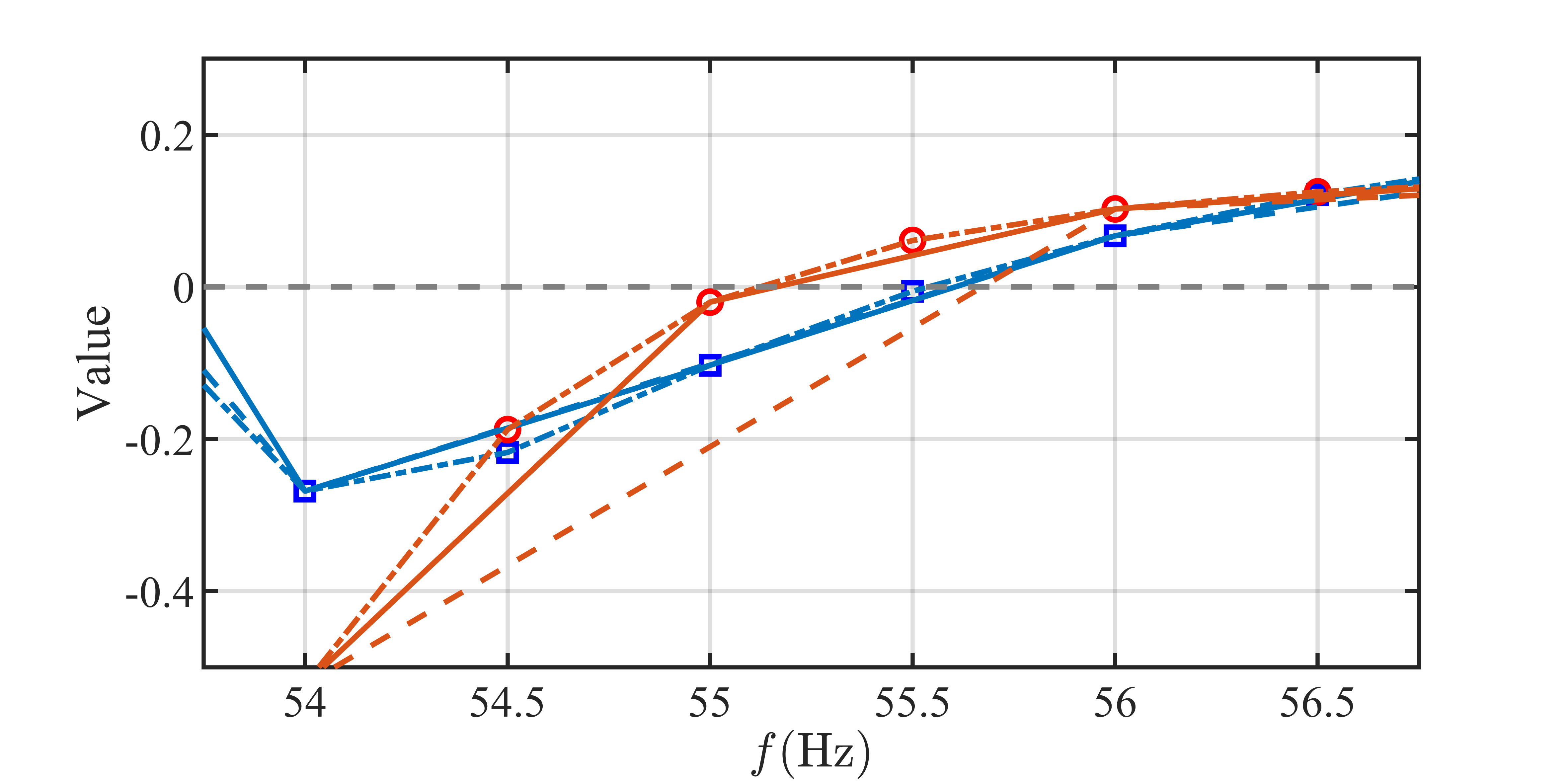} 
		\label{f_jiange010}
	}
	\DeclareGraphicsExtensions.
	\caption{DT curves at different frequency intervals. Dash-dotted line: 0.5Hz; Solid line: 1Hz; Dashed line: 2Hz. Circle and square: the frequency points. Blue line: real part; Red line: imaginary part.}
	\label{Fre_jiange}
\end{figure}

In order to study the effect of FS intervals on the critical pole estimation errors, different frequency intervals are explored and compared in the cases of $ K_{rp}=0.15$  and $ K_{rp}=0.1$. Set the frequency interval as 2Hz, 1Hz, and 0.5Hz respectively, and the swept DT curves near the critical pole are shown in Fig. \ref{Fre_jiange}. TABLE \ref{Table_f_jiange} gives the calculated critical poles with different frequency intervals.

When the frequency interval is large (2Hz), there is a significant error between the estimated and the reference poles. In the case of  $K_{rp}=0.1$, the real part of the estimated critical pole is even opposite to that of the reference, indicating that the stability assessment conclusion is wrong in this situation. If a smaller frequency interval (1Hz or 0.5Hz) is adopted, the estimated errors are significantly reduced, and all stability assessment conclusions are correct. Considering the increased workload of 0.5Hz frequency interval, the performance improvement is unworthy.

To conclude, smaller frequency intervals bring more accurate impedance models but also introduce heavier workloads. According to the simulation and experimental cases, setting a frequency interval of 1Hz for the DFIG grid-connected system is reasonable.

\emph{2) Frequency response processing}

The frequency-coupled model is a series of discrete points ${{\bf{Y}}_{{\rm{DFIG}}}}(j{\omega _1}),{{\bf{Y}}_{{\rm{DFIG}}}}(j{\omega _2}), \cdots ,{{\bf{Y}}_{{\rm{DFIG}}}}(j{\omega _n})$  after FS. According to Section II-C, the qualitative stability conclusion can be obtained by the IDTA diagram, which is directly computed by discrete data of $ {{\bf{Y}}_{{\rm{DFIG}}}} $. Nevertheless, a continuous frequency model is required for the quantitative analysis, especially within the neighborhood of the critical poles. Thus, the piecewise linear interpolation algorithm is introduced for its computational efficiency. 

In the experimental case IV-B, when $ C_s=800 $$ \mu $F  and $ C_s=600 $$ \mu $F, the data near the critical pole are processed by the piecewise linear interpolation algorithm, and Fig. \ref{fitting} shows the results. Other data processing methods, including the cubic polynomial fitting and Lagrange interpolation algorithms, are also explored for comparison. The critical pole can be calculated based on the obtained data according to (\ref{damping}). TABLE \ref{tab_com_do} lists the critical poles obtained by different methods. The interval of FS is 1Hz, and the step size of the interpolation/fitting algorithm is 0.1Hz. The actual critical poles computed by experimental data are also provided as the baseline.

\begin{table}[t]
	\begin{center}
		\caption{Critical poles calculated from the data of different frequency intervals}
		\label{Table_f_jiange}
		\begin{tabular}{ c | c |  c  }
			\toprule
			Frequency interval	 & $K_{rp}=0.15$ &  $K_{rp}=0.1$ \\
			\midrule
			2Hz &-0.103 + 350.789i &-0.302 + 349.802i \\
			\midrule
			1Hz &-0.163+348.894i &0.259+345.991i \\
			\midrule
			0.5Hz & -0.176+ 348.782i &  0.261 + 346.844i \\
			\midrule
			Reference	& -0.167+348.818i & 0.261+345.987i \\
			\bottomrule
		\end{tabular}
	\end{center}
\end{table}	

TABLE \ref{tab_com_do} shows that the Lagrange interpolation has the least errors, followed by the cubic polynomial fitting and finally the piecewise linear interpolation algorithm. Overall, all three frequency response processing algorithms have acceptable accuracy, especially in the imaginary part. Considering the significant computation burden of Lagrange interpolation and the mediocre performance of the cubic polynomial fitting algorithm, the piecewise linear interpolation is recommended to estimate the critical poles of the grey-box DFIG system.

\subsection{Application Potential}
The proposed APSAM is convenient for engineering applications, and it is simple and intuitive to utilize the IDTA diagram for stability evaluation. Besides, when the system is stable, the method can further conduct quantitative stability analysis by estimating the critical poles of the system. Estimating critical poles can reflect the stability margin of the grey-box system and then provide guidance for stability control.

Similar to other small-signal stability assessment methods, the proposed method can apply the stability analysis of DFIG at single operating points, but the stability margin information provided by the proposed method helps avoid the heavy FS and stability assessment workload at all operating points. Specific application scenarios include:

\begin{figure}[t]
	\centering
	\subfigure[$ C_s $=800$ \mu $F]{
		\includegraphics[width=0.8\linewidth]{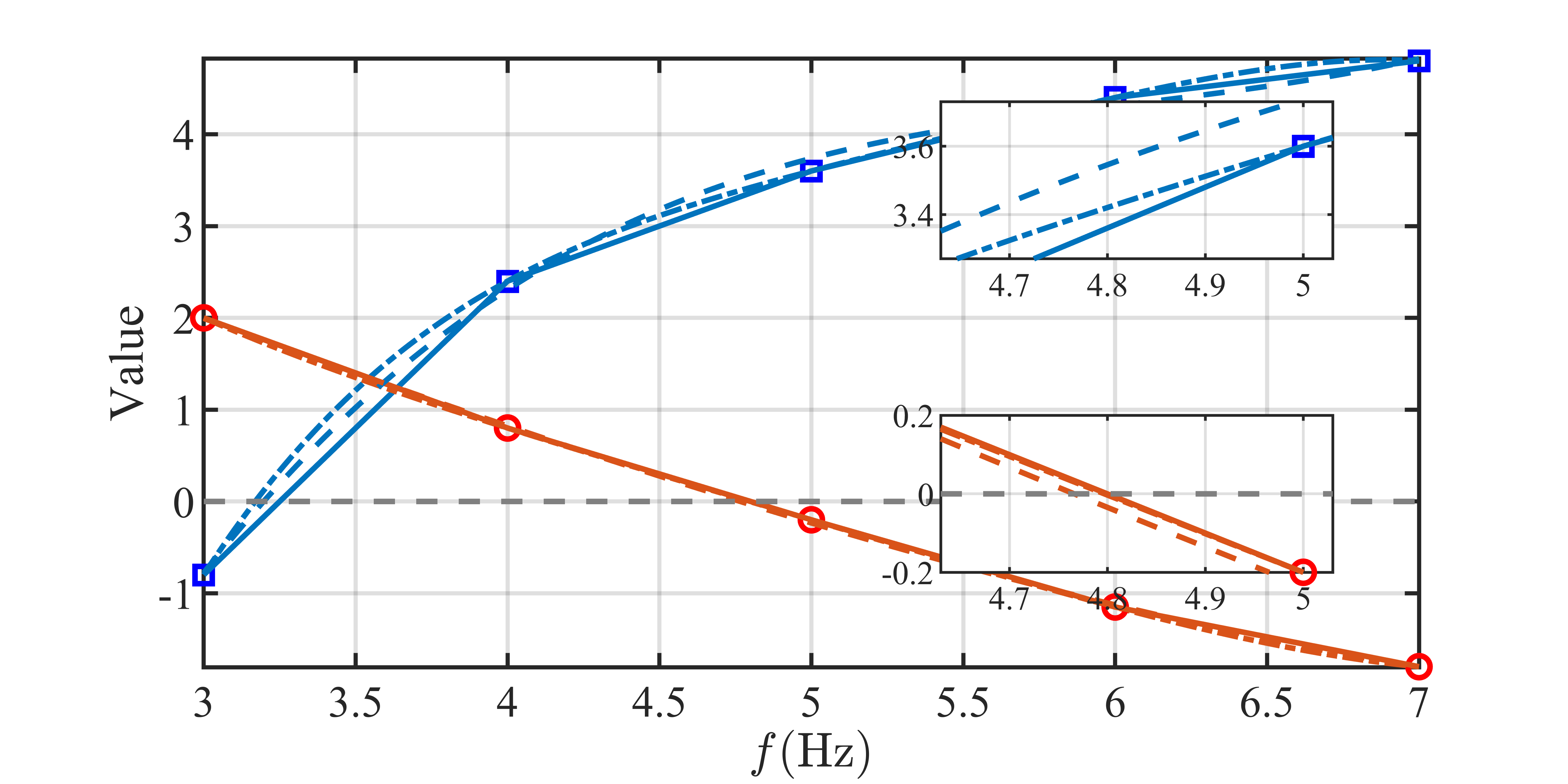}
		\label{fitting_800}
	}
	\subfigure[$ C_s $=600$ \mu $F]{
		\includegraphics[width=0.8\linewidth]{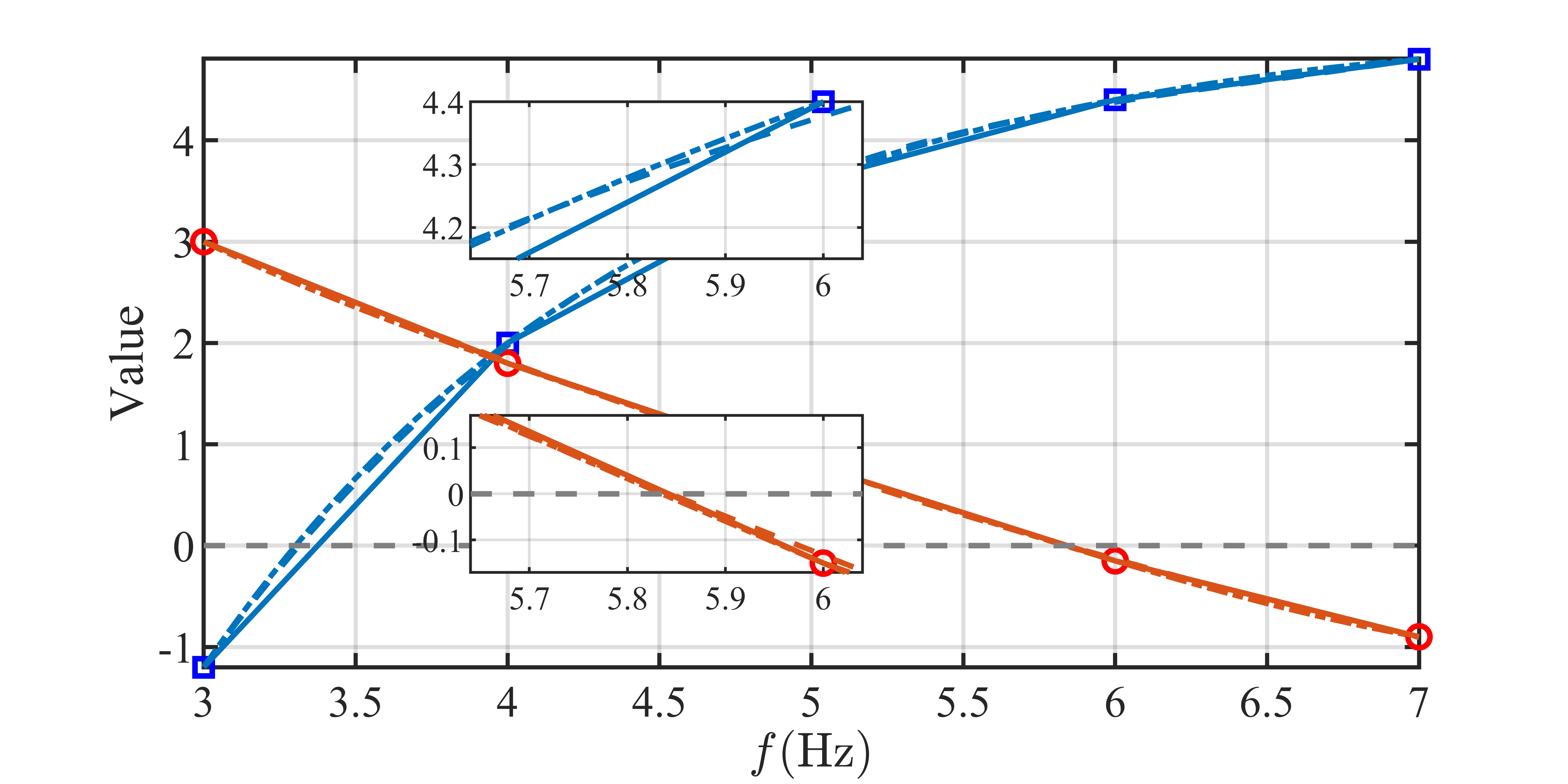} 
		\label{fitting_600}
	}
	\DeclareGraphicsExtensions.
	\caption{Data processing under different $ C_{s} $. Square and circle: discrete data; Solid line: piecewise linear interpolation; Dashed line: cubic polynomial fitting; Dash-dotted line: Lagrange interpolation; Grey dashed line: zero line.}
	\label{fitting}
\end{figure}

\begin{table}[t]
	\begin{center}
		\caption{Comparison of the calculated critical pole values}
		\label{tab_com_do}
		\begin{tabular}{ c | c  c  c  }
			\toprule
			$ C_s $	 & 800$ \mu $F & 600$ \mu $F & \thead{Error\\(Real \& Imag)}\\
			\midrule
			\thead{Experimental Data} &2.174+28.903i &3.226+32.673i & - \\
			\midrule
			\textbf{\thead{Piecewise linear \\ interpolation}} &1.456+28.475i &2.747+34.579i & 0.599 \& 1.167 \\
			\midrule
			\thead{Cubic poly-\\nomial fitting} &1.604+28.224i &3.557+34.468i &  0.451 \& 1.237\\
			\midrule
			\thead{Lagrange \\interpolation} &1.815+28.398i &3.194+34.501i& 0.196 \&  1.166\\
			\bottomrule
		\end{tabular}
	\end{center}
\end{table}
	
\begin{figure}[t]
	\centering
	\includegraphics[width=1.00\linewidth]{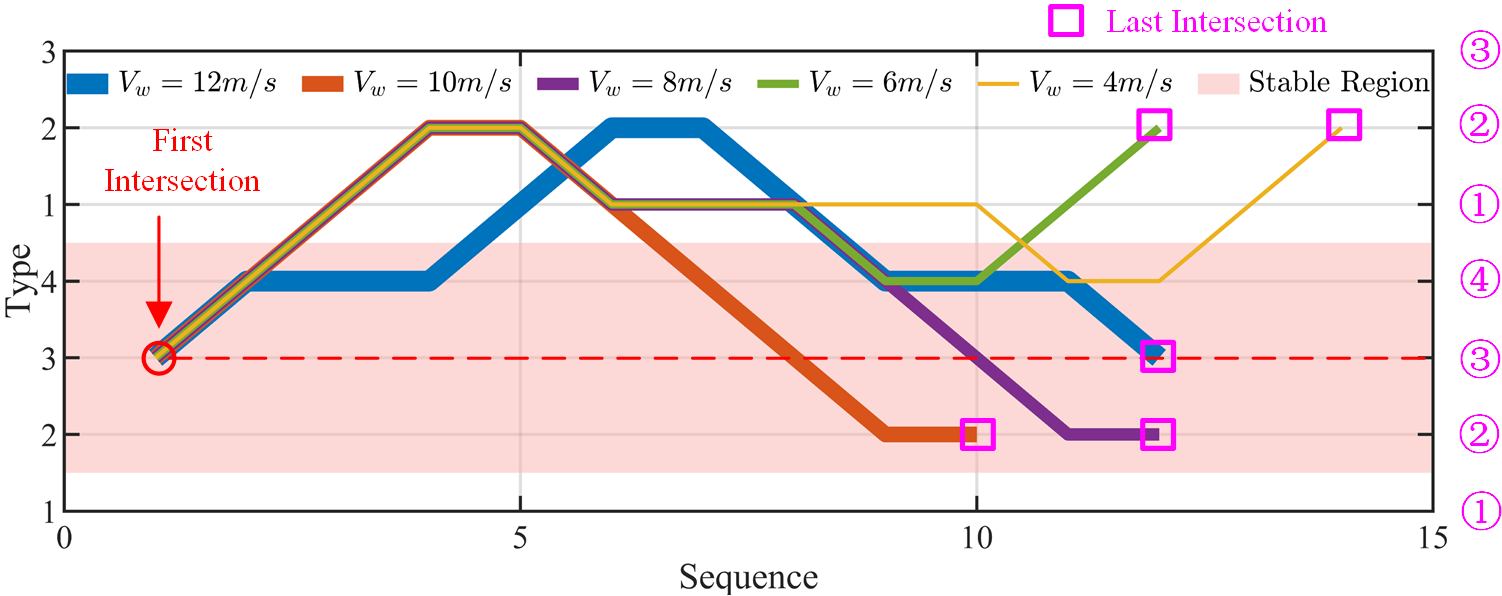}
	\caption{IDTA diagrams under different wind speeds. }
	\label{dif_Vw1}
\end{figure}

1) If offline stability analysis is required for the grey-box system, only the operating point with the worst stability needs to be analyzed \cite{Yang_PLL}. For the DFIG grid-tied system, low wind speed usually indicates worse small-signal stability \cite{zukang3_3},\cite{NC7}.  This is corroborated in Fig. \ref{dif_Vw1}, which displays the IDTA diagrams of the DFIG system at different wind speeds. The results indicate that the scenarios with high speed (12m/s, 10m/s, 8m/s) are stable, while that with low speed (6m/s and 4m/s) are unstable. Under such situations, only the critical scenarios (6m/s and 8m/s) need to be analyzed according to the critical poles’ estimation. Thus, the stability analysis on high-stability margin scenarios (10m/s, 12m/s) can be avoided. 
	
2) Stability analysis can be performed online if an online FS device \cite{online_scan} is available. At this time, the critical poles calculated by APSAM can reflect the system stability at different moments, and the method is also applicable even for the systems considering controller parameter limits \cite{Li_parameter_limit}. The preventative control can be activated when the system stability margin is minor. 
	
Moreover, the system stability can be judged directly according to the IDTA diagram. In Fig. \ref{dif_Vw1}, the stability of different scenarios can be determined by checking
whether the last intersection of the IDTA curve falls into the stable region. The entire stability determination process is automated and intuitive, facilitating its engineering applications. 

\section{Conclusion}
In this paper, an Argument-principle based stability assessment method (APSAM) for grey-box MIMO systems is proposed. The results of time-domain simulations and hardware-in-loop experiments show that the proposed APSAM can accurately evaluate the system's stability using the black-box device's frequency response. When applied in practice, the system's stability can be judged intuitively according to the IDTA diagram. Moreover, the method can evaluate the system's critical pole to provide a quantitative stability assessment. The control of the grey-box DFIG system will be researched based on the APSAM in the near future.
\setcounter{table}{0}  
\renewcommand{\thetable}{A\Roman{table}}
\setcounter{equation}{0}  
\renewcommand{\theequation}{A\arabic{equation}}
\setcounter{figure}{0}  
\renewcommand{\thefigure}{A\arabic{figure}}
\appendix
\subsection{A Stability Misjudgment Case}
\label{stability_mis}
\begin{figure}[t]
	\centering
	\includegraphics[width=0.8\linewidth]{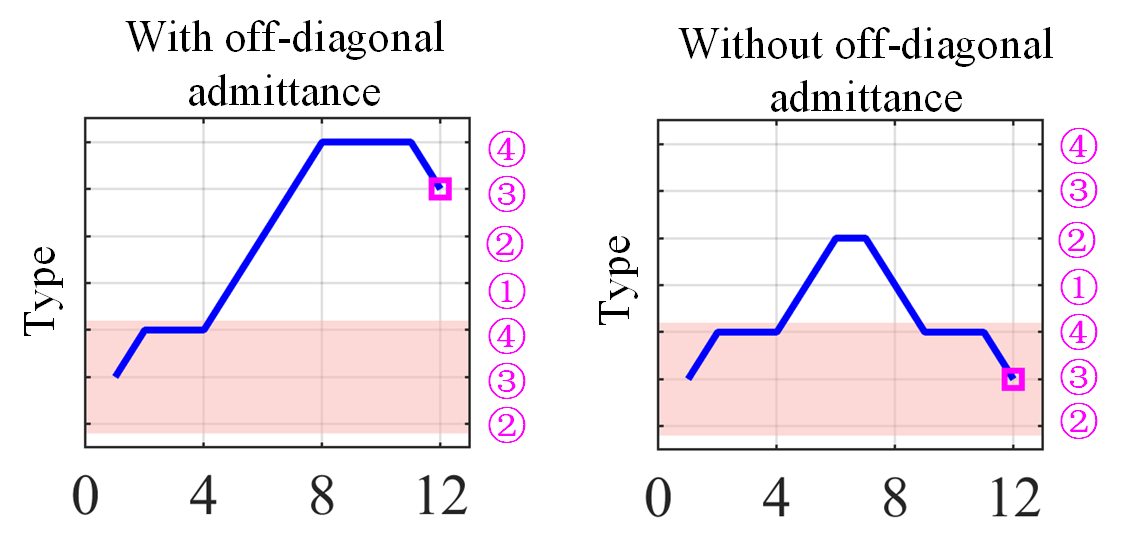}
	\caption{IDTA diagrams considering off-diagonal admittance or not.  Square: last intersection; Shadow: stable region. }
	\label{mis_IDTA}
\end{figure}
\begin{figure}[t]
	\centering
	\subfigure[With off-diagonal admittance.]{
		\includegraphics[width=1\linewidth]{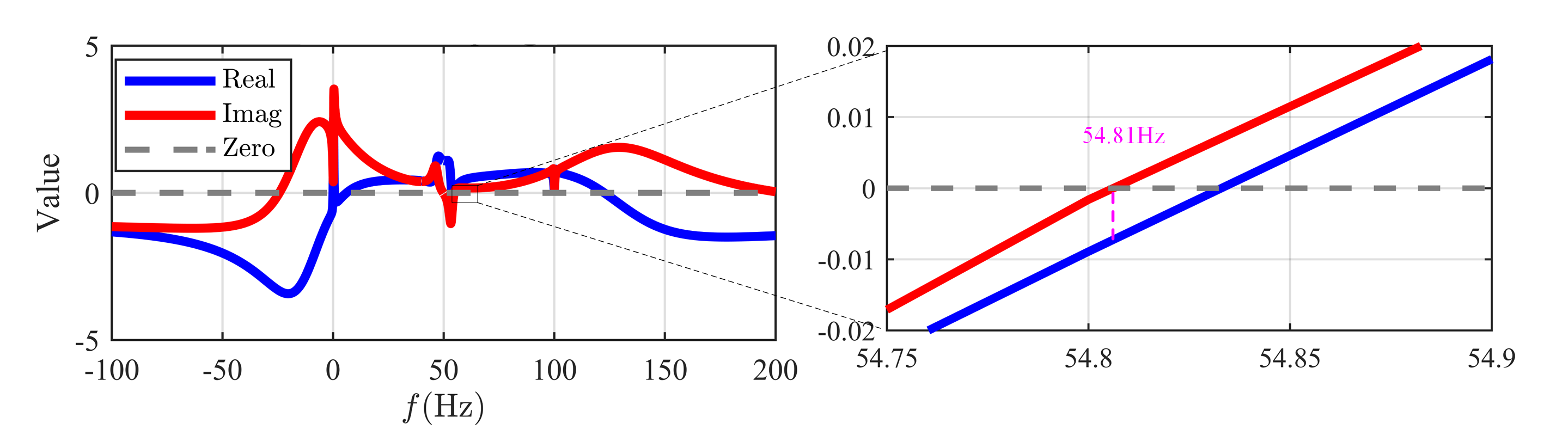}
		\label{with_cou}
	}
	\subfigure[Without off-diagonal admittance.]{
		\includegraphics[width=1\linewidth]{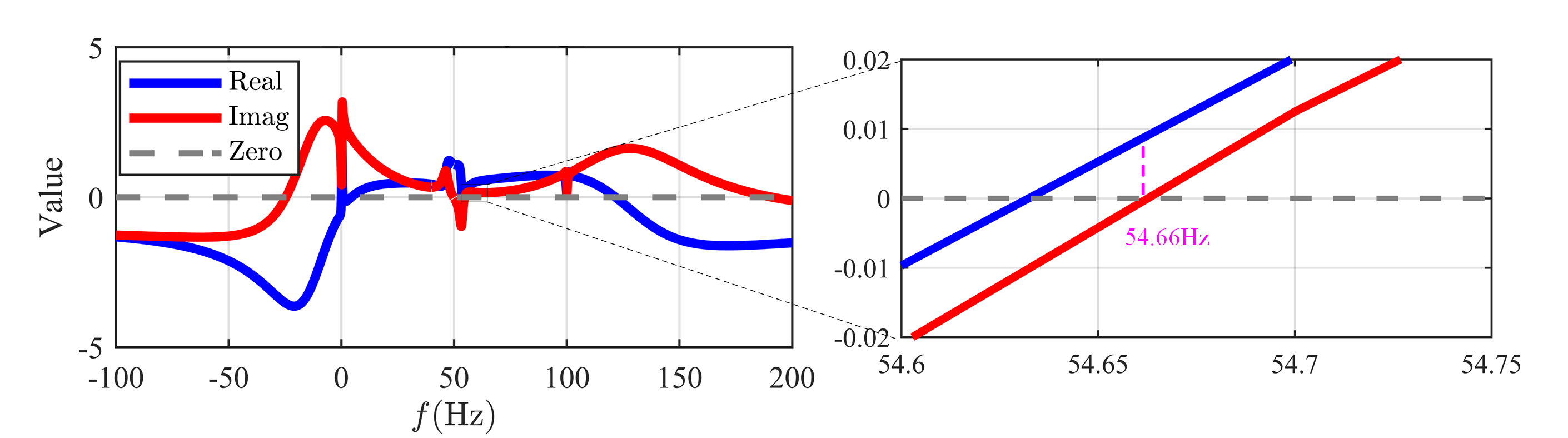} 
		\label{without_cou}	
	}
	\DeclareGraphicsExtensions.
	\caption{DT curves considering off-diagonal admittance or not.}
	\label{mis_DT}
\end{figure}
\begin{table}[b]
	\renewcommand\arraystretch{1.2}
	\begin{center}
		\caption{Calculated critical pole considering off-diagonal admittance or not.}
		\label{mis_cri_z}
		\begin{tabular}{ c | c  c  c c c }
			\hline
			Situation	 & $ f $ & Re[$ D(s) $] & $ a $ & $ b $ &$ z_o $ of $ D(s) $\\
			\hline
			1& 54.81Hz &-0.0073 &0.282&-0.275&0.013+344.4i\\
			2& 54.66Hz &0.0083 &0.328&-0.300&-0.014+343.4i\\
			\hline
		\end{tabular}
	\end{center}
\end{table}
\begin{figure}[t]
	\centering
	\subfigure[DT curves in the key frequency band. Red: imaginary curve; Blue: Real curve.]{
		\includegraphics[width=0.85\linewidth]{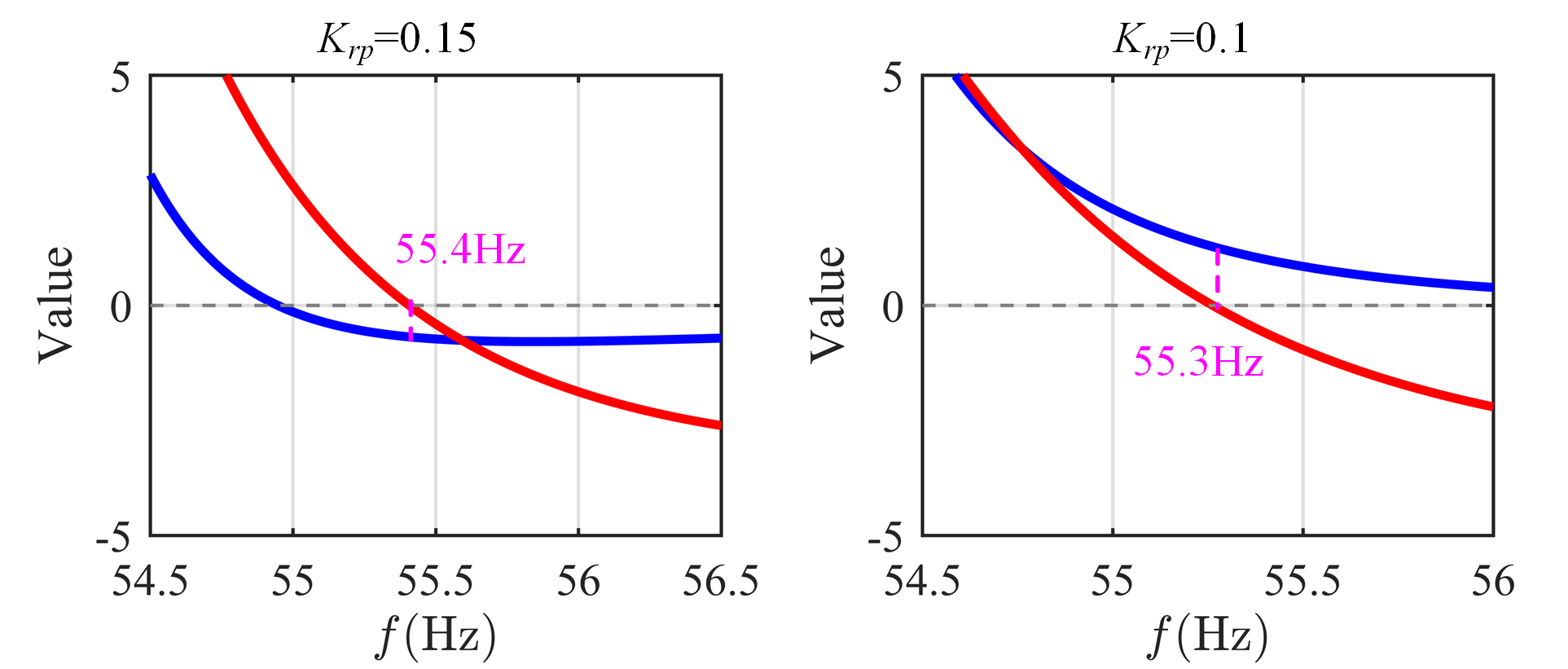}
		\label{curve015_010}
	}
	\subfigure[IDTA diagrams. Square: last intersection; Shadow: stable region.]{
		\includegraphics[width=0.85\linewidth]{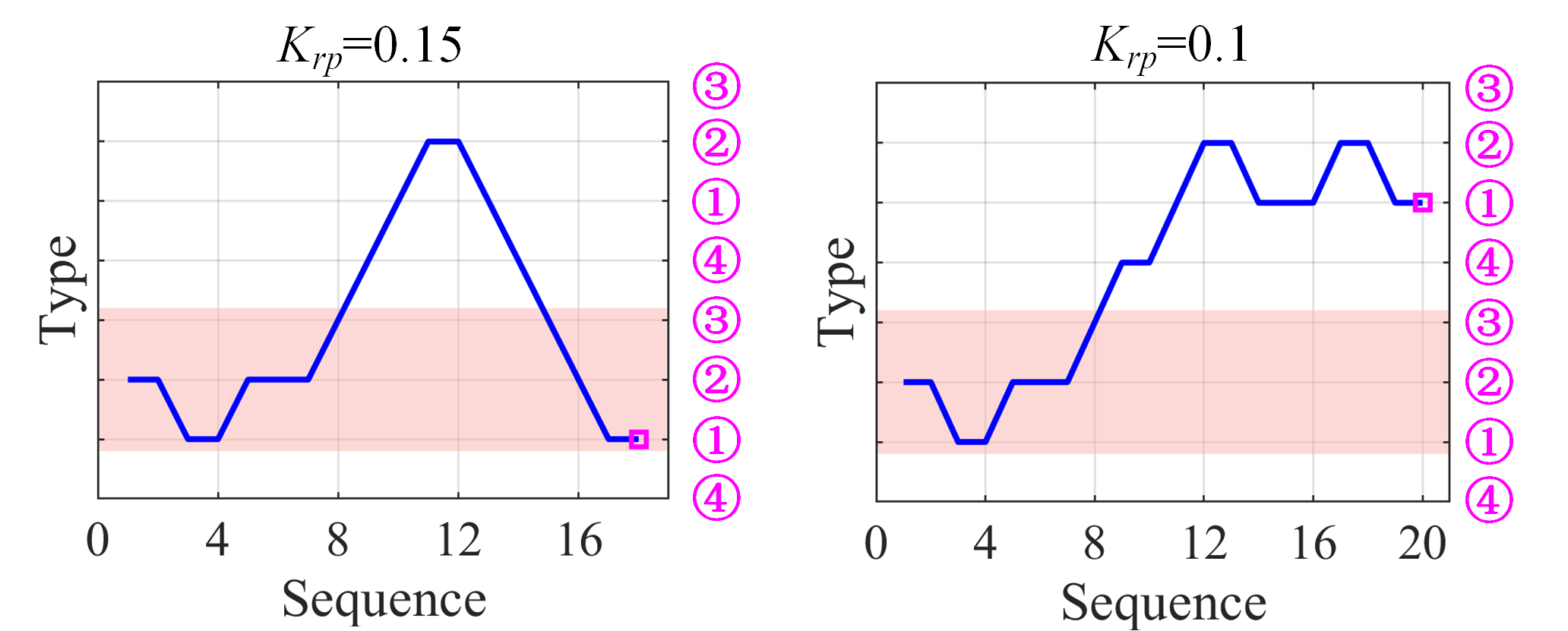} 
		\label{IDTA015_010}	
	}
	\DeclareGraphicsExtensions.
	\caption{The trajectory characteristics of $ D(s) $ based on the DFIG's impedance under different $ K_{rp} $.}
	\label{imp}
\end{figure}

For the target DFIG in \ref{sim_case}, the d- and q-axis parameters in the control loops may be different, which will increase the frequency coupling effect of DFIG \cite{NC6}. Let $ K_{rp,d}=0.15 $ and $ K_{rp,q}=0.1 $, Fig. \ref{mis_IDTA} represents the IDTA diagrams considering the off-diagonal admittance or not. It can be shown that the last intersection falls outside the stable region when considering the off-diagonal admittance; whereas the last intersection falls into the stable region when ignoring the off-diagonal admittance.	
Fig. \ref{mis_DT} displays the DT curves of the system considering the DFIG's off-diagonal admittance or not. It can be seen that the intersection types near 55Hz are inconsistent in two situations. The corresponding calculated critical poles ($ z_o $ of $ D(s) $) are displayed in TABLE \ref{mis_cri_z}, where Situation 1 takes into account off-diagonal admittance whereas Situation 2 does not. It is clear that the real part of the system's critical poles is opposite in the two situations. This means that the stability results obtained in the two situations are opposite. Therefore, ignoring off-diagonal admittance may lead to stability misjudgment.

\subsection{Consistency Verification}
\label{model_selec}
The impedance model of the DFIG is employed to analyze the system stability for the simulated instance in \ref{sim_case}. Let 
\begin{equation}
D(s) = \det ({\bf{I}} + {{\bf{Z}}_{{\rm{DFIG }}}}(s){{\bf{Y}}_{\rm{grid}}}(s)).
	\label{im_D}
\end{equation} 
Then the target system’s stability under $ {K_{rp}} = 0.15 $ and $ {K_{rp}} = 0.1 $ can be analyzed using the proposed APSAM. The $ D(s) $ trajectories in the key frequency band are shown in Fig. A\ref{curve015_010}, and Fig. A\ref{IDTA015_010} presents the IDTA diagrams.

When $ {K_{rp}} = 0.15 $, the last intersection falls in the stable region, indicating that the system is stable. While $ {K_{rp}} = 0.1 $, the last intersection falls outside the stable region, meaning that the system is unstable. The calculated critical poles of the target system based on Fig. A\ref{curve015_010} are shown in TABLE \ref{imp_cri_z}. Compared with the calculated parameters in TABLE \ref{tab_omega_damp_simulation}, it can be found that the critical poles ($ z_o $ of $ D(s) $) calculated by the DFIG's admittance and the impedance are in agreement.
\begin{table}[t]
	\renewcommand\arraystretch{1.2}
	\begin{center}
	\caption{Calculated critical pole based on the DFIG's impedance.}
		\label{imp_cri_z}
		\begin{tabular}{ c | c  c  c c c }
			\hline
			$ K_{rp} $	 & $ f $ & Re[$ D(s) $] & $ a $ & $ b $ &$ z_o $ of $ D(s) $\\
			\hline
			0.15& 55.4Hz &-0.6812 &-4.7088&0.6000&-0.1430+348.08i\\
			0.1& 55.3Hz &1.1945 &-4.5035&2.1280&0.2518+347.60i\\
			\hline
		\end{tabular}
	\end{center}
\end{table}

Combined with the analysis in \ref{sim_case}, it elucidates that for the DFIG grid-connected system, the stability analysis results derived using DFIG's admittance model and the impedance model are consistent.

\begin{IEEEbiography}[{\includegraphics[width=1in,height=1.25in,clip,keepaspectratio]{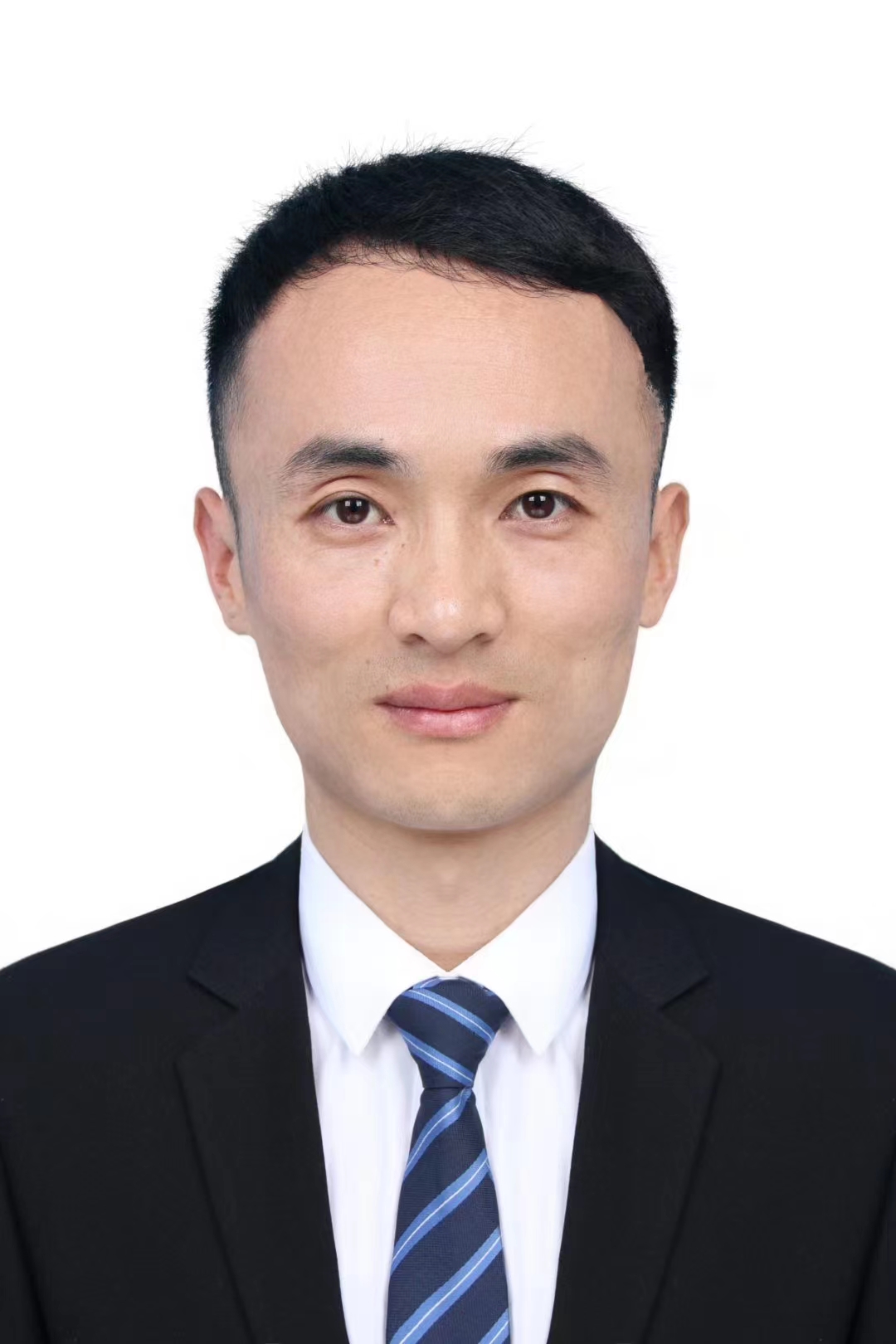}}]{Tao Zhang}
	(S’20) received the B.S. degree in electrical engineering from Xi’an Jiaotong University, Xi’an, China in 2017. And he is currently pursuing a Ph.D. degree on electrical engineering in Xi’an Jiaotong University.  
	
	His research interest includes small-signal stability analysis and control of renewable energy power systems.
\end{IEEEbiography}
\begin{IEEEbiography}[{\includegraphics[width=1in,height=1.25in,clip,keepaspectratio]{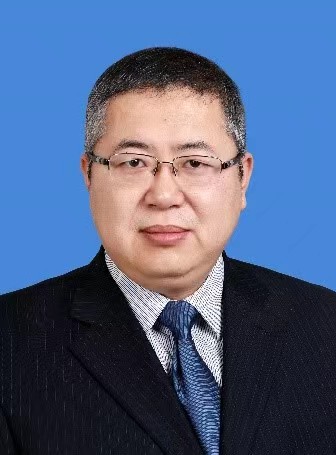}}]{Zhiguo Hao}
	(M’10) was born in Ordos, China, in 1976. He received his B.S. and Ph.D. degrees in electrical engineering from Xi'an Jiaotong University, Xi'an, China, in 1998 and 2007, respectively. 
	
	He has been a Professor with the Electrical Engineering Department, Xi'an Jiaotong University since 2018. His research interest includes protection and control of power systems and equipment.
\end{IEEEbiography}
\begin{IEEEbiography}[{\includegraphics[width=1in,height=1.25in,clip,keepaspectratio]{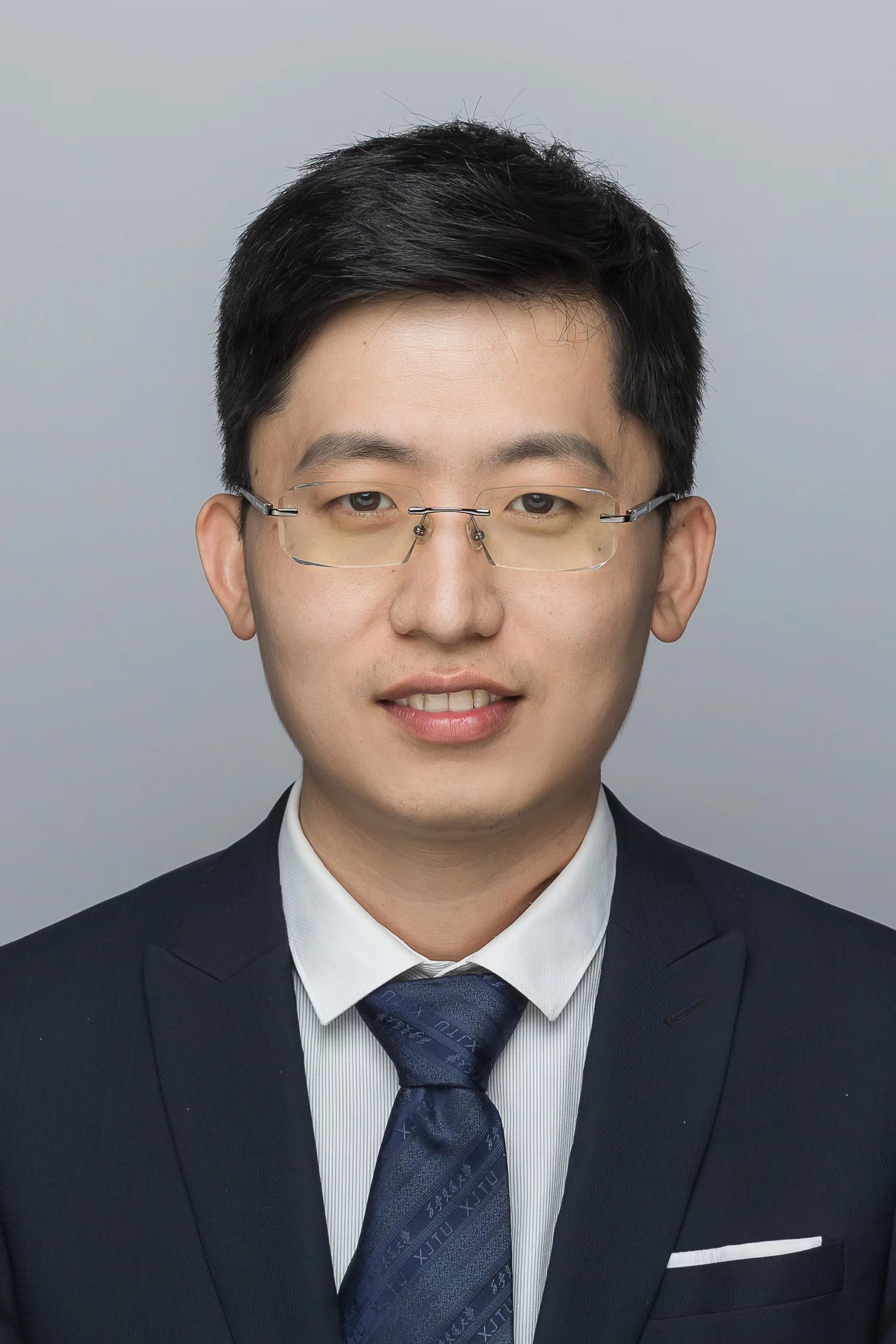}}]{Songhao Yang}
(S’18-M’19) was born in Shandong, China, in 1989. He received the B.S. and Ph.D. degrees in electrical engineering from the Xi’an Jiaotong University, Xi’an, China, in 2012 and 2019, respectively. Besides, he received the Ph.D. degree in electrical and electronic engineering from Tokushima University, Japan, in 2019. 
	
	Currently, he is an Associate Professor at Xi’an Jiaotong University. His research interest includes power system stability analysis and control.
\end{IEEEbiography}
\begin{IEEEbiography}[{\includegraphics[width=1in,height=1.25in,clip,keepaspectratio]{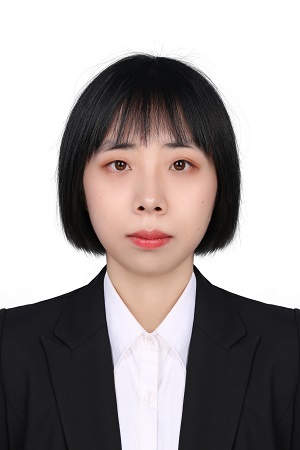}}]{Hongyue Ma}
received the B.S. degree from Harbin University of Science and Technology, Harbin, China in 2015, and M.S. degree from Xi’an Jiaotong University, Xi’an, China in 2020, where she is working toward a Ph.D degree, all in electrical engineering. 

 Her research interest includes renewable energy power system analysis and protection.
\end{IEEEbiography}
\begin{IEEEbiography}[{\includegraphics[width=1in,height=1.25in,clip,keepaspectratio]{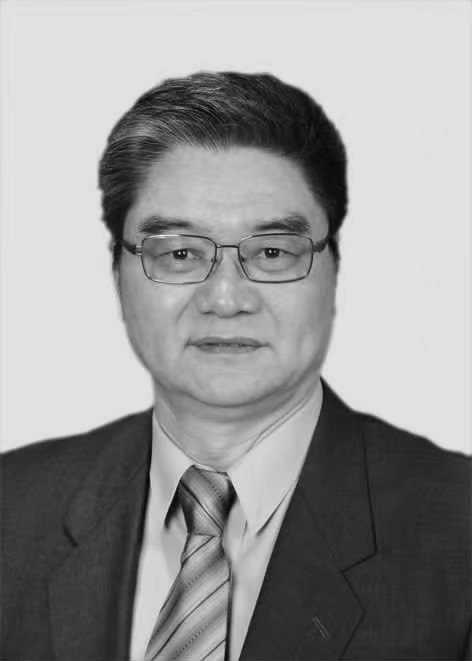}}]{Baohui Zhang}
	 (SM’99-F’19) was born in Hebei Province, China, in 1953. He received the M.Eng. and Ph.D. degrees in electrical engineering from Xi’an Jiaotong University, Xi’an, China, in 1982 and 1988, respectively. 
	 
	 He has been a Professor in the Electrical Engineering Department at Xi’an Jiaotong University since 1992. His research interests are power system analysis, control, communication, and protection.
\end{IEEEbiography}

\end{document}